\documentclass[twocolumn]{aastex631}
\hypersetup{linkcolor=blue,urlcolor=blue}
\usepackage{amsmath}
\usepackage{CJKutf8}

\shorttitle{The Environment of the Earliest Luminous Quasars}
\shortauthors{Wang et al.}

\begin{document}

\title{ASPIRE: The Environments and Dark Matter Halos of Luminous Quasars in the Epoch of Reionization}
\suppressAffiliations 

\correspondingauthor{Feige Wang}
\email{fgwang@umich.edu}

\author[0000-0002-7633-431X]{Feige Wang}
\affiliation{Department of Astronomy, University of Michigan, 1085 S. University Ave., Ann Arbor, MI 48109, USA}

\author[0000-0002-6184-9097]{Jaclyn B. Champagne}
\affiliation{Steward Observatory, University of Arizona,933 N Cherry Ave, Tucson, AZ 85721}
\affiliation{Space Telescope Science Institute, 3700 San Martin Dr, Baltimore, MD 21218, USA}
\author[0000-0002-5721-0709]{Jiamu Huang}
\affiliation{Department of Physics, University of California, Santa Barbara, CA 93106-9530, USA}
\author[0000-0001-5287-4242]{Jinyi Yang}
\affiliation{Department of Astronomy, University of Michigan, 1085 S. University Ave., Ann Arbor, MI 48109, USA}
\author[0000-0002-7054-4332]{Joseph F.\ Hennawi}
\affiliation{Department of Physics, University of California, Santa Barbara, CA 93106-9530, USA}
\affiliation{Leiden Observatory, Leiden University, Niels Bohrweg 2, NL-2333 CA Leiden, Netherlands}
\author[0000-0003-3310-0131]{Xiaohui Fan}
\affiliation{Steward Observatory, University of Arizona, 933 N Cherry Avenue, Tucson, AZ 85721, USA}
\author[0000-0002-4321-3538]{Haowen Zhang}
\affiliation{Steward Observatory, University of Arizona, 933 N Cherry Avenue, Tucson, AZ 85721, USA}
\author[0000-0002-6748-2900]{Tiago Costa}
\affiliation{School of Mathematics, Statistics and Physics, Newcastle University, Newcastle upon Tyne, NE1 7RU, UK}
\author[0000-0002-2662-8803]{Roberto  Decarli}
\affiliation{INAF–Osservatorio di Astrofisica e Scienza dello Spazio, via Gobetti 93/3, I-40129, Bologna, Italy}
\author{Melanie Habouzit}
\affiliation{Department of Astronomy, University of Geneva, Chemin Pegasi 51, Versoix CH-1290, Switzerland}
\author[0000-0002-4622-6617]{Fengwu Sun}
\affiliation{Center for Astrophysics $\vert$\ Harvard\ \&\ Smithsonian, 60 Garden St., Cambridge, MA 02138, USA}
\author[0000-0002-2931-7824]{Eduardo Ba\~nados}
\affiliation{Max Planck Institut f\"ur Astronomie, K\"onigstuhl 17, D-69117 Heidelberg, Germany}
\author[0000-0002-5768-738X]{Xiangyu Jin}
\affiliation{Department of Astronomy, University of Michigan, 1085 S. University Ave., Ann Arbor, MI 48109, USA}
\affiliation{Steward Observatory, University of Arizona, 933 N Cherry Avenue, Tucson, AZ 85721, USA}
\author[0000-0001-6874-1321]{Koki Kakiichi}
\affiliation{Cosmic Dawn Center (DAWN), Denmark}
\affiliation{Niels Bohr Institute, University of Copenhagen, Jagtvej 128, 2200 Copenhagen N, Denmark}
\author[0000-0001-5492-4522]{Romain A. Meyer}
\affiliation{Department of Astronomy, University of Geneva, Chemin Pegasi 51, 1290 Versoix, Switzerland}
\author[0000-0003-0111-8249]{Yunjing Wu}
\affiliation{Department of Astronomy, Tsinghua University, Beijing 100084, China}
\author{Silvia Belladitta}
\affiliation{Max Planck Institut f\"ur Astronomie, K\"onigstuhl 17, D-69117 Heidelberg, Germany}
\affiliation{INAF–Osservatorio di Astrofisica e Scienza dello Spazio, via Gobetti 93/3, I-40129, Bologna, Italy}
\author[0000-0002-2183-1087]{Laura Blecha}
\affiliation{Department of Physics, University of Florida, Gainesville, FL, 32611, USA}
\author[0000-0001-8582-7012]{Sarah E.~I.~Bosman}
\affiliation{Institute for Theoretical Physics, Heidelberg University, Philosophenweg 12, D–69120, Heidelberg, Germany}
\affiliation{Max Planck Institut f\"ur Astronomie, K\"onigstuhl 17, D-69117 Heidelberg, Germany}
\author[0000-0001-8467-6478]{Zheng Cai}
\affiliation{Department of Astronomy, Tsinghua University, Beijing 100084, China}
\author[0000-0002-7898-7664]{Thomas Connor}
\affiliation{Center for Astrophysics $\vert$\ Harvard\ \&\ Smithsonian, 60 Garden St., Cambridge, MA 02138, USA}
\author[0000-0003-0821-3644]{Frederick B.\ Davies}
\affiliation{Max Planck Institut f\"ur Astronomie, K\"onigstuhl 17, D-69117 Heidelberg, Germany}
\author[0000-0003-2895-6218]{Anna-Christina Eilers}
\affiliation{Department of Physics, Massachusetts Institute of Technology, Cambridge, MA 02139, USA}
\affiliation{MIT Kavli Institute for Astrophysics and Space Research, Massachusetts Institute of Technology, Cambridge, MA 02139, USA}
\author[0000-0003-3633-5403]{Zolt\'an Haiman}
\affiliation{Institute of Science and Technology Austria (ISTA), Am Campus 1, Klosterneuburg 3400, Austria}
\affiliation{Department of Astronomy, Columbia University, New York, NY 10027, USA}
\affiliation{Department of Physics, Columbia University, New York, NY 10027, USA}
\author[0000-0003-1470-5901,gname='Hyunsung',sname='Jun']{Hyunsung D. Jun}
\affiliation{Department of Physics, Northwestern College, 101 7th St SW, Orange City, IA 51041, USA}
\affiliation{School of Physics, Korea Institute for Advanced Study, 85 Hoegiro, Dongdaemun-gu, Seoul 02455, Republic of Korea}
\author[0000-0001-6251-649X]{Mingyu Li}
\affiliation{Department of Astronomy, Tsinghua University, Beijing 100084, China}
\author[0000-0001-5951-459X]{Zihao Li}
\affiliation{Cosmic Dawn Center (DAWN), Denmark}
\affiliation{Niels Bohr Institute, University of Copenhagen, Jagtvej 128, 2200 Copenhagen N, Denmark}
\author[0000-0003-3762-7344]{Weizhe Liu \begin{CJK}{UTF8}{gbsn}(刘伟哲)\end{CJK}}
\affiliation{Steward Observatory, University of Arizona, 933 N Cherry Avenue, Tucson, AZ 85721, USA}
\author{Alessandro Lupi}
\affiliation{Como Lake Center for Astrophysiscs, DiSAT, Universit\'a degli Studi dell'Insubria, via Valleggio 11, 22100, Como, Italy}
\affiliation{INFN, Sezione di Milano-Bicocca, Piazza della Scienza 3, 20126 Milano, Italy}
\author[0000-0002-6221-1829]{Jianwei Lyu (\begin{CJK}{UTF8}{gbsn}吕建伟\end{CJK})}
\affiliation{Steward Observatory, University of Arizona, 933 N Cherry Avenue, Tucson, AZ 85721, USA}
\author[0000-0002-5941-5214]{Chiara Mazzucchelli}
\affiliation{Instituto de Estudios Astrof\'{\i}sicos, Facultad de Ingenier\'{\i}a y Ciencias, Universidad Diego Portales, Avenida Ejercito Libertador 441, Santiago, Chile}
\author[0000-0003-2984-6803]{Masafusa Onoue}
\affiliation{Kavli Institute for Astronomy and Astrophysics, Peking University, Beijing 100871, China}
\affiliation{Kavli Institute for the Physics and Mathematics of the Universe (Kavli IPMU, WPI), The University of Tokyo, Chiba 277-8583, Japan}
\author[0000-0002-9712-0038]{Elia Pizzati}
\affiliation{Leiden Observatory, Leiden University, Niels Bohrweg 2, NL-2333 CA Leiden, Netherlands}
\affiliation{Center for Astrophysics $\vert$\ Harvard\ \&\ Smithsonian, 60 Garden St., Cambridge, MA 02138, USA}
\author[0000-0003-4924-5941]{Maria Pudoka}
\affiliation{Steward Observatory, University of Arizona, 933 N Cherry Avenue, Tucson, AZ 85721, USA}
\author[0000-0003-2349-9310]{Sof\'ia Rojas-Ruiz}\affiliation{Department of Physics and Astronomy, University of California, Los Angeles, 430 Portola Plaza, Los Angeles, CA 90095, USA}
\author[0000-0002-4544-8242]{Jan-Torge Schindler}
\affiliation{Hamburg Observatory, University of Hamburg, Gojenbergsweg 112, D-21029 Hamburg, Germany}
\author[0000-0003-1659-7035]{Yue Shen}
\affiliation{Department of Astronomy, University of Illinois Urbana-Champaign, Urbana, IL 61801, USA}
\affiliation{National Center for Supercomputing Applications, University of Illinois Urbana-Champaign, Urbana, IL 61801, USA}
\author[0000-0003-0747-1780]{Wei Leong Tee}
\affiliation{Department of Astronomy and Astrophysics, The Pennsylvania State University, 525 Davey Lab, University Park, PA 16802, USA}
\affiliation{Institute for Gravitation and the Cosmos, The Pennsylvania State University, University Park, PA 16802, USA}
\author[0000-0002-3683-7297]{Benny Trakhtenbrot}
\affiliation{School of Physics and Astronomy, Tel Aviv University, Tel Aviv 69978, Israel}
\affiliation{Max-Planck-Institut f{\"u}r extraterrestrische Physik, Gie\ss{}enbachstra\ss{}e 1, 85748 Garching, Germany}
\affiliation{Excellence Cluster ORIGINS, Boltzmannsstra\ss{}e 2, 85748, Garching, Germany}
\author[0000-0002-6849-5375]{Maxime Trebitsch}
\affil{LUX, Observatoire de Paris, Universit\'e PSL, Sorbonne Universit\'e, CNRS, 75014 Paris, France}
\author[0000-0001-9191-9837]{Marianne Vestergaard}
\affiliation{DARK, Niels Bohr Institute, The University of  Copenhagen, Jagtvej  155, DK-2200  Copenhagen N, Denmark}
\affiliation{Steward Observatory, University of Arizona, 933 N Cherry Avenue, Tucson, AZ 85721, USA}
\author[0000-0002-3216-1322]{Marta Volonteri}
\affiliation{Institut d’Astrophysique de Paris, UMR 7095, CNRS and Sorbonne Universit\'e, 98 bis Boulevard Arago, 75014 Paris, France}
\author[0000-0003-4793-7880]{Fabian Walter}
\affiliation{Max Planck Institut f\"ur Astronomie, K\"onigstuhl 17, D-69117 Heidelberg, Germany}
\author[0000-0002-0123-9246]{Huanian Zhang} 
\affiliation{Department of Astronomy, Huazhong University of Science and Technology, Wuhan, Hubei 430074, China}
\author[0000-0002-3983-6484]{Siwei Zou}
\affiliation{Chinese Academy of Sciences South America Center for Astronomy, National Astronomical Observatories, CAS, Beijing 100101, China}
\affiliation{Departamento de Astronom\'ia, Universidad de Chile, Casilla 36-D, Santiago, Chile}

\begin{abstract}
We present a systematic study of the environments of 25 luminous quasars at $z > 6.5$ from the ASPIRE program. Using JWST/NIRCam WFSS data, we identified 487 galaxies at $5.3 \lesssim z \lesssim 7.0$ exhibiting [\ion{O}{3}] emission. Among these, 122 [\ion{O}{3}] emitters lie within $|\Delta v_{\rm los}| < 1000~\mathrm{km~s^{-1}}$ of the quasars, corresponding to a $\sim9.4$-fold enhancement relative to the average galaxy density at other redshifts. Furthermore, we identified 16 [\ion{C}{2}]-emitting galaxies at the quasar redshifts from ALMA mosaic observations. A cross-correlation function (CCF) analysis between quasars and [\ion{O}{3}]+[\ion{C}{2}] emitters yields a cross-correlation length of $r_0^{\rm QG} = 8.68^{+0.51}_{-0.55}~h^{-1}~\mathrm{cMpc}$ and a auto-correlation of $r_0^{\rm{QQ}}=15.76^{+2.48}_{-2.70}~h^{-1}~{\rm cMpc}$, indicating that $z \sim 7$ quasars reside in dark matter halos with $M_{\rm halo} = 10^{12.27^{+0.21}_{-0.26}}~M_\odot$ and has a quasar lifetime of $t_{\rm Q}=\rm 10^{{7.05}^{+0.95}_{-1.01}}~yr$. Notably, the number of [\ion{O}{3}]-emitting galaxies at quasar redshifts varies significantly from field to field, ranging from zero to twenty, highlighting a diverse quasar environment. Remarkably, seven quasars trace significant galaxy overdensities (i.e., protoclusters), with $\delta_{\rm gal} > 5$ within a volume of $V \sim 500~{\rm cMpc^3}$. We also find that $|\Delta v_{\rm los}|$ increases rapidly toward smaller galaxy-quasar separations in protocluster fields, consistent with galaxy kinematics around extremely massive halos in cosmological simulations. By combining JWST and ALMA data, we reveal the complex and diverse environments of these early quasars, providing robust evidence that the earliest luminous quasars are effective tracers of galaxy overdensities, albeit with substantial field-to-field variation.
\end{abstract}

\keywords{Early universe (435) --- Galaxies(573) --- Protoclusters (1297) --- Redshift surveys (1378) --- Supermassive black holes (1663)}


\section{Introduction} \label{sec:intro}
Understanding the formation of the earliest supermassive black holes (SMBHs) has been one of the frontiers of extragalactic astronomy. Since the discovery of the first $z > 6$ quasar \citep{Fan01}, over two decades of painstaking searches have yielded a sample of $\sim$50 luminous quasars with absolute magnitudes of $M_{\rm 1450}\lesssim -25$ at $z > 6.5$, extending up to the current redshift record at $z = 7.6$ \citep{Wang21a}. The existence of these quasars powered by $\sim10^9$ $M_\odot$ black holes \citep[e.g.,][and references therein]{Banados18a, Yang20a, Yang21, Farina22} during the Epoch of Reionization (EoR) challenges our understanding of supermassive black hole (SMBH) formation,  {and implies that these quasars must have grown rapidly within the first few hundred million years after the Big Bang. Such rapid growth requires either the formation of massive seed black holes, such as those formed via direct collapse \citep{Begelman06} or through dense stellar clusters \citep{Portegies04} , or sustained periods of accretion at or above the Eddington limit from smaller seeds \citep{Madau01}. However, such conditions are difficult to achieve and maintain, given the expected low gas densities, strong radiative feedback from the central SMBHs, and short available timescales in the early Universe. Together with the low space density \citep[e.g.,][]{Wang19b} and large black hole masses of luminous quasars \citep{Yang21, Farina22} , most theoretical works favor the idea that the most distant luminous quasars reside in massive dark matter halos and can be traced by large-scale galaxy overdensities \citep{Costa14,DiMatteo17}. Such environments can provide mechanisms that suppress cooling and fragmentation, enabling the formation of massive black hole seeds, and offer sufficient gas supplies to sustain continuous black hole growth \citep[see][for a review]{Inayoshi20}.}

Indeed, an assortment of cosmological simulation models can produce such massive SMBHs \citep[e.g.,][]{DiMatteo05,Khandai15,DiMatteo17} powering quasars with roughly the observed abundance $0.4~{\rm Gpc}^{-3}$ \citep{Wang19b} starting with massive $\gtrsim 10^4~M_\odot$ seed BHs. These models generically predict that these SMBHs are hosted by massive $M_\star \gtrsim 10^{11}M_\odot$ galaxies and reside in the rarest $M_{\rm DM} > 10^{12}M_\odot$ dark matter halos situated in the most overdense regions in the early Universe \citep{Costa14}. While these numerical works have established the plausibility of the existence of $z\sim7$ quasars in a cosmological context, to date, rigorously testing these theories remains highly challenging. Specifically, the following two critical questions need to be resolved observationally in order to test these theoretical works: 
(1) In which dark matter halos and large-scale environments do these SMBHs live?
(2) Do the earliest supermassive black holes coevolve with their host galaxies?

If distant quasars inhabit massive dark-matter halos, they are expected to be good tracers of galaxy overdensities or protoclusters in the early Universe \citep{Overzier22}. Extensive studies have been made in the past two decades for searching for galaxy overdensities around $z\gtrsim6$ quasars, however, no conclusive results have emerged from HST or ground-based observations \citep[e.g.,][]{Kim09, Simpson14, Mignoli20, Pudoka24}. 
{On the other hand, recent ALMA [\ion{C}{2}] surveys have revealed key insights into the environments of high-redshift quasars \citep{Decarli17, Trakhtenbrot17, Neeleman19b, Nguyen20,Venemans20, Meyer22, Wang23, Wang24a}. Notably, \citet{Meyer22} found a striking overdensity of [\ion{C}{2}] emitters around $z>6$ quasars, with number densities several orders of magnitude above that of blank-field survey results, far exceeding results from UV- or optically-bright galaxies \citep[i.e., Lyman break galaxies;][]{Pudoka24,Mignoli20}. However, these ALMA studies typically probe only the immediate vicinity ($<1$ cMpc) with single-pointing ALMA observations, limiting the ability to distinguish small-scale mergers from the large-scale environment that traces massive dark matter halos. In addition, the small field of view of ALMA and the shallow depth of existing observations could introduce additional geometrical and luminosity biases, affecting our ability to interpret the scatter in the number of detected satellite galaxies \citep{Zana22, Zana23}.} 
More recently, infrared surveys enabled by the wide-field slitless spectroscopy (WFSS) with JWST NIRCam is changing the game. The recent EIGER project enabled the first study of the clustering of galaxies around $z\sim 6$ quasars. It measured the quasar-galaxy cross-correlation function around four quasars and determined quasars' host dark matter halos to be ${\rm log}~M_{\rm halo}=12.43^{+0.13}_{-0.15}~{\rm M_\odot}$ \citep{Eilers24}. 
{\cite{Schindler26} also attempted to constrain the host dark matter halos of $z>7$ quasars using two quasar fields observed with NIRCam imaging and NIRSpec/MSA follow-up spectroscopy.} However, this measurement is significantly affected by cosmic variance due to the limited number of quasar sightlines used in these works.

The ASPIRE program, short for ``A Spectroscopic Survey of Biased Halos in the Reionization Era" (GO-2078; PI Wang), is designed to overcome the above limitations. ASPIRE is a JWST quasar legacy survey which targets a sample of 25 luminous quasars at redshift between 6.5 and 6.8. Exploiting the excellent sensitivity of NIRCam/WFSS at $\sim3-4~\mu$m, ASPIRE aims to detect a large sample of star-forming galaxies with strong [\ion{O}{3}] emission lines at $5.3\lesssim z \lesssim7$ along these quasar sight lines. In particular, ASPIRE obtains a complete sample of galaxies at the quasar redshifts down to a line flux limit of $\rm\sim2\times10^{-18}~erg~s^{-1}~cm^{-2}$. In addition, ASPIRE enables identification of galaxies from the local Universe up to $z\sim9$ using other line tracers (e.g., Pa$\alpha$, [\ion{S}{3}], \ion{He}{1}, Pa$\beta$, [\ion{O}{2}]). Furthermore, ASPIRE includes coordinated NIRISS parallel imaging observations, which detects additional galaxies photometrically. Finally, ASPIRE is accompanied by an ALMA large program (Program ID:2022.1.01077.L; PI: Wang) which performs mosaic observations around these 25 quasars at 1mm. The ALMA observations are designed for detecting [\ion{C}{2}] emitting galaxies at the quasar redshifts and dusty star-forming galaxies at both low- and high-redshifts. 

A series of studies using ASPIRE data has already been published. For instance, \cite{Wang23} reported the discovery of a protocluster around one ASPIRE quasar; \cite{Yang23} analyzed the rest-frame optical spectral properties of a sample of ASPIRE quasars; \cite{Wu23} and \cite{Zou24} investigated the host galaxies of some metal absorbers and their circumgalactic medium (CGM) properties; \cite{Lin24} reported a sample of broad-line H$\alpha$ emitters in the early Universe; \cite{Jin24} and \cite{Kakiichi25} found direct evidence of early reionization around [\ion{O}{3}] emitting galaxies; \cite{Champagne24a, Champagne24b} studied the structure of the protocluster discovered by \cite{Wang23} and environment-dependent galaxy evolution; \cite{Sun25} measured the obscured cosmic star formation rate density using a spectroscopically complete sample of galaxies at $z \simeq 4-6$; and \cite{Li25a, Li25b} studied the metal enrichment history of galaxies from the local universe up to redshift $z\sim9$.

In this paper, we provide an overview of the ASPIRE program, present data reduction procedures of JWST NIRCam and ALMA observations, report the discovery of 487 star-forming galaxies with [\ion{O}{3}] emission lines at $5<z<7$ and 14 galaxies with  [\ion{C}{2}] emission lines at the quasar redshifts, and provide a detailed study of the environment and the dark matter halo mass of the ASPIRE quasars by combing JWST and ALMA observations.  
Throughout the paper, we adopt a $\Lambda$CDM cosmology with $H_0=70~{\rm km~s^{-1}~Mpc^{-1}}$, $\rm \Omega_M=0.3$, and $\Omega_\Lambda=0.7$.
In this paper, we organize as follows:
we present an overview of the ASPIRE program and the data used for this work in \S \ref{sec:obs}, 
show the [\ion{O}{3}] emitting galaxy searching algorithms and report the discovery of 487 [\ion{O}{3}] emitting galaxies in \S \ref{sec:emitters}, we also report the discovery of the [\ion{C}{2}] emitting galaxies found at quasar redshifts in \S \ref{sec:emitters}, 
present clustering analyses for both [\ion{O}{3}] emitting galaxies and [\ion{C}{2}] emitting galaxies in \S \ref{sec:clustering} and investigate the diversity of quasar environments and discuss implementations in \S \ref{sec:environment}, 
 and summarize in \S \ref{sec:summary}. 

\begin{figure}
\centering
\includegraphics[width=0.49\textwidth]{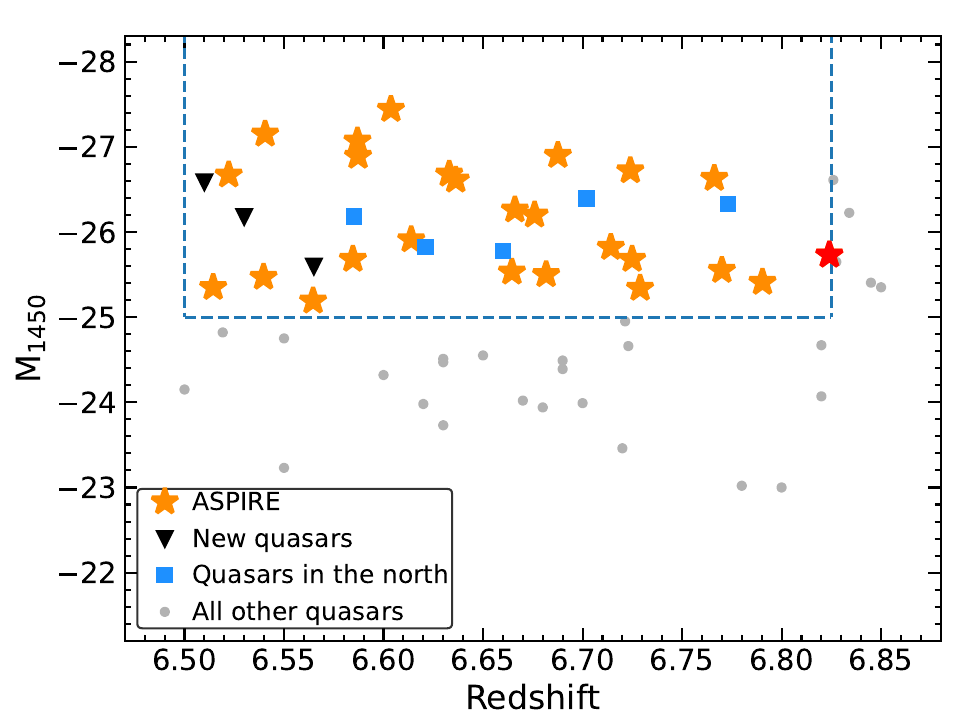}
\caption{{\bf The redshift and absolute magnitude distribution of ASPIRE quasars.}
We selected a flux-limited sample of 25 quasars at $6.5<z\lesssim6.8$ with $M_{1450}<-25.0$ (orange asterisks). To ensure all ASPIRE quasars can be observed with ALMA, we excluded known quasars with $\rm Decl. > 30^\circ$. This sample also includes a radio-loud quasar (red asterisk, $z=6.82$). The triangles denote three quasars satisfy our luminosity cut but were published after JWST Cycle 1 proposal deadline and therefore were not included in ASPIRE.  
\label{fig:sample}}
\end{figure}

\begin{deluxetable*}{ccccccccccccccc}
\tablecaption{ASPIRE Quasar Sample.}\label{tbl:targets}
\setlength{\tabcolsep}{2pt}
\tabletypesize{\scriptsize}
\tablehead{\colhead{ID} & \colhead{$z$} & \colhead{$M_{1450}$} & \colhead{$L_{\rm bol}$} & \colhead{$M_{\rm BH}$} & \colhead{Disc. Ref.} & \colhead{$z$ Ref.} & \colhead{Ref.} & \colhead{Obs. ID} }
\startdata
             &        &          & $\rm 10^{46}~erg~s^{-1}$            & $\rm 10^9~M_\odot$ & & & \\
(1)          &  (2)   & (3)      &            (4)&            (5)& (6)                   & (7)                   & (8)           & (9) \\
\hline
J0109$-$3047 & 6.7904 & $-$25.41 &  7.69$\pm$0.20& 1.11$\pm$0.40 & \cite{Venemans13}     & \cite{Venemans20}     & \cite{Farina22} & 1 \\
J0218$+$0007 & 6.7700 & $-$25.55 &  6.40$\pm$1.40& 0.61$\pm$0.07 & \cite{Yang21}         & \cite{Wang24b}        & \cite{Yang21}    & 2 \\
J0224$-$4711 & 6.5222 & $-$26.67 &   33.6$\pm$2.0& 1.30$\pm$0.18 & \cite{Reed17}         & \cite{Wang24b}        & \cite{Wang21b}  & 3 \\
J0226$+$0302 & 6.5405 & $-$27.15 & 24.96$\pm$0.15& 3.74$\pm$0.59 & \cite{Venemans15a}    & \cite{Venemans20}     & \cite{Farina22} & 4 \\
J0229$-$0808 & 6.7249 & $-$25.68 & 10.47$\pm$0.24& 0.38$\pm$0.02 & \cite{Belladitta25}       & \cite{Wang24b}        & \cite{Belladitta25}    & 5 \\
J0244$-$5008 & 6.7240 & $-$26.72 &   14.4$\pm$0.2& 1.15$\pm$0.39 & \cite{Reed19}         & \cite{Reed19}         & \cite{Reed19}   & 6 \\
J0305$-$3150 & 6.6139 & $-$25.91 &  9.42$\pm$0.12& 0.54$\pm$0.12 & \cite{Venemans19}     & \cite{Venemans20}     & \cite{Farina22} & 7 \\
J0430$-$1445 & 6.7142 & $-$25.82 & 10.23$\pm$0.24& 1.29$\pm$0.03 & \cite{Belladitta25}          & \cite{Wang24b}        & \cite{Belladitta25}   & 8 \\
J0525$-$2406 & 6.5397 & $-$25.47 &    6.8$\pm$3.5& 0.29$\pm$0.04 & \cite{Yang21}         & \cite{Wang24b}        & \cite{Yang21}   & 9 \\
J0706$+$2921 & 6.6037 & $-$27.44 &   33.9$\pm$1.5& 2.11$\pm$0.16 & \cite{Wang19b}        & \cite{Wang24b}        & \cite{Yang21}   & 10 \\
J0910$+$1656 & 6.7289 & $-$25.34 &    5.3$\pm$0.6& 0.41$\pm$0.03 & \cite{Wang19b}        & \cite{Wang24b}        & \cite{Yang21}   & 111\\
J0910$-$0414 & 6.6363 & $-$26.61 &   15.0$\pm$1.1& 3.59$\pm$0.61 & \cite{Wang19b}        & \cite{Wang24b}        & \cite{Yang21}   & 12 \\
J0921$+$0007 & 6.5646 & $-$25.19 &    6.1$\pm$0.6& 0.26$\pm$0.01 & \cite{Yang21}         & \cite{Wang24b}        & \cite{Yang21}   & 13 \\
J0923$+$0402 & 6.6330 & $-$26.68 &   21.7$\pm$3.0& 1.77$\pm$0.02 & \cite{Wang19b}        & \cite{Wang24b}        & \cite{Yang21}   & 14 \\
J0923$+$0753 & 6.6817 & $-$25.50 &    4.9$\pm$2.0& 0.49$\pm$0.15 & \cite{Yang21}         & \cite{Wang24b}        & \cite{Yang21}   & 15 \\
J1048$-$0109 & 6.6759 & $-$26.20 & 10.55$\pm$0.30& 2.29$\pm$0.64 & \cite{Wang17}         & \cite{Venemans20}     & \cite{Farina22} & 16 \\
J1058$+$2930 & 6.5846 & $-$25.68 &    5.8$\pm$1.5& 0.54$\pm$0.03 & \cite{Yang21}         & \cite{Wang24b}        & \cite{Yang21}   & 17 \\
J1104$+$2134 & 6.7662 & $-$26.63 &   15.1$\pm$0.9& 1.69$\pm$0.15 & \cite{Wang19b}        & \cite{Wang24b}        & \cite{Yang21}   & 18 \\
J1110$-$1329 & 6.5144 & $-$25.35 &    5.5$\pm$0.6& 0.38$\pm$0.14 & \cite{Venemans15a}    & \cite{Venemans20}     & \cite{Wang21b}  & 19 \\
J1129$+$1846 & 6.8240 & $-$25.73 &    8.4$\pm$1.9& 0.29$\pm$0.02 & \cite{Banados21}      & \cite{Yang21}         & \cite{Yang21}   & 20 \\
J1526$-$2050 & 6.5869 & $-$27.07 & 22.34$\pm$0.54& 4.06$\pm$1.00 & \cite{Mazzucchelli17} & \cite{Venemans20}     & \cite{Farina22} & 71 \\
J2002$-$3013 & 6.6876 & $-$26.90 &   15.4$\pm$1.9& 1.62$\pm$0.27 & \cite{Yang21}         & \cite{Wang24b}        & \cite{Yang21}   & 22 \\
J2102$-$1458 & 6.6645 & $-$25.53 &    6.0$\pm$0.5& 0.74$\pm$0.11 & \cite{Wang19b}        & \cite{Wang24b}        & \cite{Yang21}   & 23 \\
J2132$+$1217 & 6.5872 & $-$26.89 & 20.04$\pm$0.21& 1.13$\pm$0.14 & \cite{Mazzucchelli17} & \cite{Venemans20}     & \cite{Farina22} & 124  \\
J2232$+$2930 & 6.6660 & $-$26.26 &   10.0$\pm$1.7& 3.06$\pm$0.36 & \cite{Venemans15a}    & \cite{Mazzucchelli17} & \cite{Yang21} & 25 
\enddata
\tablenotetext{}{
(1): ID for ASPIRE quasar fields determined from quasar coordinates (i.e., JHHMM+/-DDMM).
(2): The most accurate redshift measured for each quasar, mostly determined from [\ion{C}{2}] observations.
(3): The absolute magnitude at rest-frame 1450 $\rm \AA$ of each quasar.
(4): The bolometric luminosity of each quasar.
(5): The black hole mass of each quasar.
(6): References for quasar discoveries.
(7): References for quasar redshifts.
(8): References for $M_{1450}$, bolometric luminosity and black hole mass.
(9): JWST observation ID in ASPIRE program.
}
\end{deluxetable*}

\section{ASPIRE Program Design and Observations}\label{sec:obs}

\subsection{Overview}
As discussed in \S \ref{sec:intro}, one of the key science drivers of the ASPIRE program is to probe the large-scale environment and the dark matter halos of the earliest quasars. A fundamental result of the $\rm \Lambda$CDM  structure formation paradigm is that the clustering of a population can be directly related to their host dark halo masses \citep{Mo02}. However, the extremely low density of $z\simeq7$ quasars precludes the possibility of measuring their auto-correlation function. A complementary approach for determining quasar host halo masses is to measure the clustering of galaxies around them. If quasars and galaxies trace the same dark matter overdensities, the quasar-galaxy cross-correlation $\xi_{\rm QG}(r)$, can be uniquely predicted given the respective auto-correlation of quasars $\xi_{\rm QQ}$ and galaxies $\xi_{\rm GG}$ \citep[e.g.,][]{Garcia17}. The key requirement on measuring the quasar-galaxy cross-correlation at high-redshift is to spectroscopically identify faint galaxies around quasars. Therefore, a spectroscopic redshift survey around a statistical sample of distant quasars is critically needed. Recent observations of high-$z$ galaxies suggest that young (1--10 Myr) stellar populations produce strong H$\beta$+[\ion{O}{3}] lines \citep[e.g.][]{Endsley21}. This motivated us observing quasars at $6.5<z\lesssim6.8$, putting them at the sweet-spot of NIRCam/WFSS sensitivity (with the F356W filter) to the H$\beta$+[\ion{O}{3}] of physically associated galaxies. In addition, such redshift cuts enables us finding galaxies associated to the full path-length of the Ly$\alpha$ and Ly$\beta$ transmission spikes and identifying host galaxies of metal absorbers at $z>5.3$ (with H$\beta$+[\ion{O}{3}] lines) and at $z>4$ (with the H$\alpha$ line). 

\begin{figure*}
\centering
\includegraphics[trim=20 30 20 30, clip, width=0.99\textwidth]{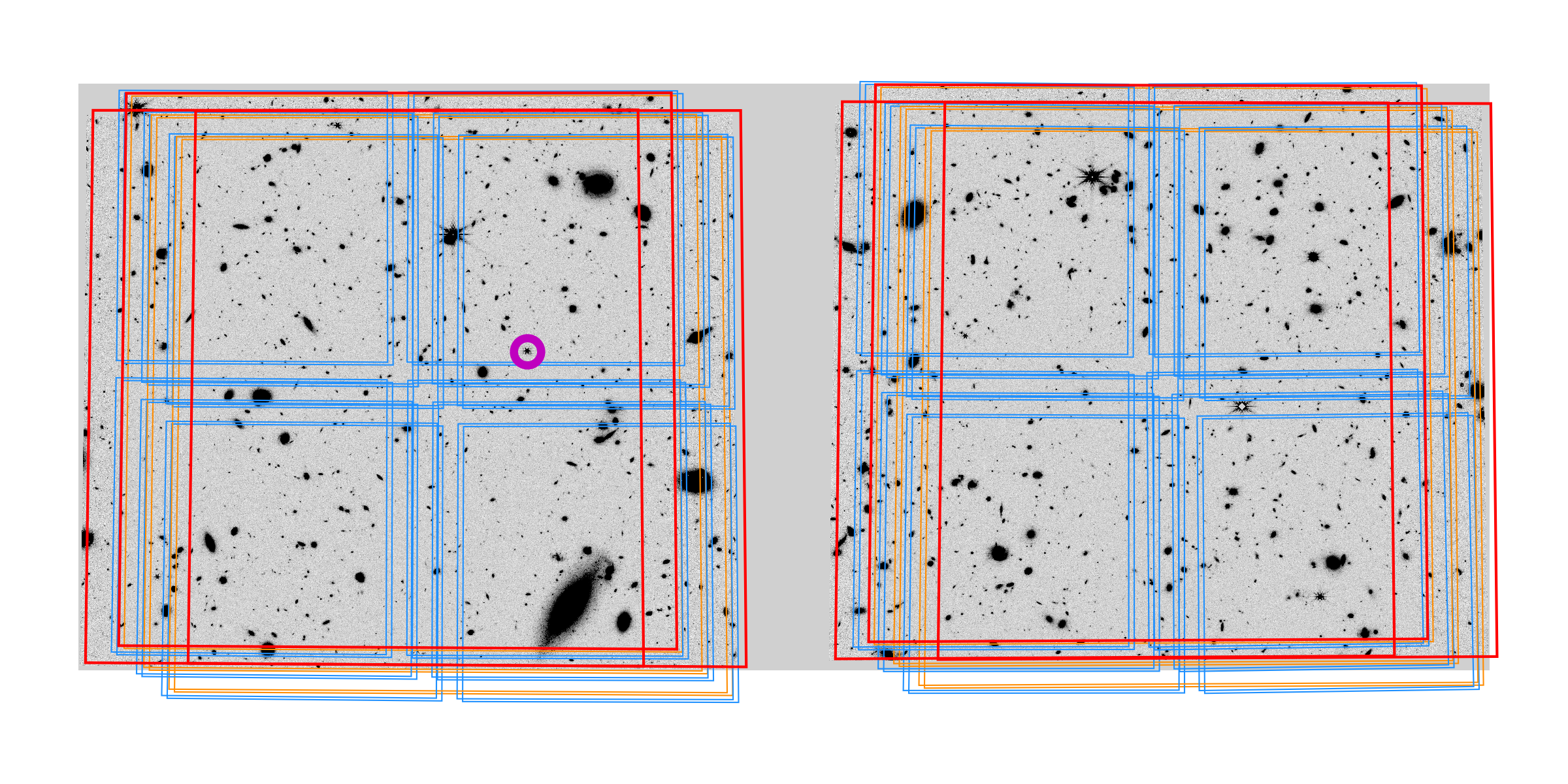}
\caption{{\bf ASPIRE NIRCam dither pattern.}
We used 3-point {\tt INTRAMODULEX} primary dither pattern with two sub-pixel dithers at each primary positions for ASPIRE observations except for J0910--0414. The orange lines highlight the WFSS dithers, while the blue lines highlight the dithers for short wavelength channel observations with F200W filter. 
The red lines denote the direct imaging and the out-of-field imaging dithers with F356W filter. The magenta circle on module A highlights the position of the quasar J0109--3047. The background image is the calibrated F356W image of this quasar field. 
}\label{fig:dither}
\end{figure*}

The ASPIRE program aims to study the most luminous quasars, which potentially host the most extreme supermassive black holes (SMBHs) in the early Universe. To achieve this goal, we selected all quasars with $M_{\rm 1450} < -25$ at $6.5 < z \lesssim 6.8$ known at the time of the JWST Cycle~1 proposal deadline (Fig.~\ref{fig:sample}). Additionally, we only targeted quasars with declinations less than $+30^\circ$, allowing us to utilize ALMA to identify dusty galaxies that could be missed by JWST observations. These selection criteria yield a sample of 25 quasars, large enough to minimize uncertainties caused by cosmic variance. Because we did not apply any cuts based on the quasars' physical properties (except for luminosity), this selection also helps reduce potential sample selection biases in measurements of quasar dark matter halo masses. 
{Since this work considers only quasars with $M_{1450} < -25$, the conclusions drawn here are applicable exclusively to these extremely luminous objects.} Notably, the sample includes the only radio-loud quasar  at $z > 6.5$ \citep{Banados21} known at that time, highlighted in Figure~\ref{fig:sample} (red asterisk). We list the basic properties of all ASPIRE quasars in Table~\ref{tbl:targets} and provide detailed descriptions of our JWST observations in the following subsections.

\subsection{JWST observations}\label{sec:jwst_obs}
We utilize the NIRCam WFSS observation mode instead of NIRCam pre-imaging followed by NIRSpec/MSA observations for the ASPIRE program to select objects with strong emission lines. We used the F356W filter for the WFSS observation in the long wavelength (LW) channel, which provides a spectral wavelength coverage of $\sim3-4\mu$m. We simultaneously obtain F200W imaging in the short wavelength channel (SW). For most of the targets, except for J0910--0414, we choose the 3-point {\tt INTRAMODULEX} primary dither pattern to cover the detector gaps in SW. In order to improve the PSF sampling, each primary position includes two sub-pixel dithers (Figure. \ref{fig:dither}). Such a dither pattern provides a survey area of $\sim$11 arcmin$^2$. For J0910--0414, we use a slightly different 4-point {\tt INTRAMODULEX} primary dither pattern and each primary position includes two sub-pixel dithers. Each quasar was placed at a carefully designed position ($\rm X_{offset}=-60\farcs5$, $\rm Y_{offset}=7\farcs5$) in module A to ensure that we have full wavelength coverage and sufficient imaging depth for the target quasar and its vicinity (see Figure \ref{fig:dither}). We used the {\tt SHALLOW4} readout pattern with nine groups and one integration for all observations. This  configuration yields a total on-source exposure time of 3779.3s for the J0910--0414 field and 2834.5 s for all other fields.

We obtained both direct and out-of-field imaging using the F115W filter in the SW channel and the F356W filter in the LW channel. The same readout pattern as adopted for the main WFSS observations was employed. The total on-source integration times are 1417.3 s near the quasar and 472.4 s at the field edge, where the coverage is limited to a single exposure, for both the F115W and F356W bands.

In addition, ASPIRE performs NIRISS observations in parallel to expand the imaging coverage around the quasars. These observations are configured with the F356W and F444W filters. Combined with ground-based optical imaging, the NIRISS data are designed to enable the photometric selection of H$\beta$+[\ion{O}{3}] emitters at $z \gtrsim 6.5$ using the $m_{\mathrm{356W}} - m_{\mathrm{444W}}$ color \cite[e.g.,][]{Smit15}. The detailed description of the NIRISS and ground-based multi-wavelength observations, as well as their analysis, will be presented in a future publication and is beyond the scope of this work. We note that the initial observations for three ASPIRE quasar fields (J0910$+$1656, J1526$-$2050, and J2132$+$1217) failed due to guide star acquisition issues and were rescheduled for later execution. The observation IDs for all ASPIRE fields are listed in Table \ref{tbl:targets}.

\begin{deluxetable}{ccccc}
\tablehead{\colhead{Field} & \colhead{Obs. Date} & \colhead{F115W} & \colhead{F200W} & \colhead{F356W}}
\startdata
J0109$-$3047 & 2022-08-11 & 27.75 & 28.43 & 28.62 \\
J0218$+$0007 & 2022-08-10 & 27.57 & 28.22 & 28.46 \\
J0224$-$4711 & 2022-08-11 & 27.54 & 28.08 & 28.35 \\
J0226$+$0302 & 2022-08-10 & 27.64 & 28.35 & 28.58 \\
J0229$-$0808 & 2022-08-16 & 27.64 & 28.23 & 28.55 \\
J0244$-$5008 & 2022-08-11 & 27.84 & 28.35 & 28.69 \\
J0305$-$3150 & 2022-08-12 & 27.78 & 28.45 & 28.70 \\
J0430$-$1445 & 2022-09-11 & 27.68 & 28.24 & 28.56 \\
J0525$-$2406 & 2022-10-12 & 27.68 & 28.18 & 28.51 \\
J0706$+$2921 & 2022-10-18 & 27.49 & 28.09 & 28.20 \\
J0910$+$1656 & 2023-11-22 & 27.44 & 28.26 & 28.14 \\
J0910$-$0414 & 2022-11-13 & 27.51 & 28.26 & 28.41 \\  
J0921$+$0007 & 2022-11-29 & 27.65 & 28.30 & 28.58 \\
J0923$+$0402 & 2022-11-29 & 27.69 & 28.31 & 28.55 \\
J0923$+$0753 & 2023-03-30 & 27.70 & 28.26 & 28.52 \\
J1048$-$0109 & 2023-04-27 & 27.71 & 28.31 & 28.55 \\
J1058$+$2930 & 2023-04-27 & 27.77 & 28.39 & 28.76 \\
J1104$+$2134 & 2022-11-28 & 27.57 & 28.29 & 28.42 \\
J1110$-$1329 & 2023-06-18 & 27.52 & 28.19 & 28.01 \\
J1129$+$1846 & 2023-06-07 & 27.48 & 28.23 & 28.15 \\
J1526$-$2050 & 2023-03-03 & 27.52 & 28.19 & 28.35 \\
J2002$-$3013 & 2022-09-22 & 27.54 & 28.09 & 28.32 \\
J2102$-$1458 & 2022-09-23 & 27.53 & 28.13 & 28.47 \\
J2132$+$1217 & 2023-10-10 & 27.15 & 27.65 & 27.81 \\
J2232$+$2930 & 2022-08-10 & 27.76 & 28.33 & 28.72 \\
\enddata
\caption{The average 5$\sigma$ limiting magnitudes measured in 0.\arcsec{32} apertures for each ASPIRE field. These are measured by placing 1000 random apertures across each NIRCam mosaic image (including both module A and module B) and measuring the median absolute deviation of flux measured in the fixed aperture size.
\label{tbl:depth}}
\end{deluxetable}

\subsection{NIRCam imaging data reduction}\label{sec:imaging_reduction}
The NIRcam images were reduced using version 1.10.2 of the JWST Calibration Pipeline\footnote{\url{https://github.com/spacetelescope/jwst}}({\tt CALWEBB}). We used the reference files (\verb|jwst_1080.pmap|) from version 11.16.21 of the standard Calibration Reference Data System (CRDS) to calibrate our data. In addition to the standard processes, we introduced several additional steps as detailed in \cite{Wang23} and \cite{Yang23} and describe them briefly below. We modeled the {\it 1/f} noise from the background subtracted {\tt rate} files ({\tt rate} files are Stage 1 outputs) on a row-by-row and column-by-column basis for each amplifier following the algorithm proposed by \cite{Schlawin2020}. The background for the {\tt rate} files was determined iteratively after masking out detected objects in the images. The {\it 1/f} noise subtracted {\tt rate} files were then fed into the Stage 2 pipeline. 
To remove extra detector-level noise features and sky background, we subtracted a master background from individual Stage 2 outputs, where the master background for each NIRCam detector was constructed from all exposures observed at a similar time. To remove astrometric offsets between individual exposures / detectors, we aligned individual images to stars selected from GAIA DR3\footnote{For a few quasar fields that do not have enough GAIA stars, we aligned images to the DESI Legacy Imaging Surveys}. Since there are too few available reference stars in individual images (especially SW images), we first passed through all LW images for each quasar fields to Stage 3 pipeline to create an initial mosaic image. The initial mosaic image was aligned to GAIA DR3 using {\tt tweakwcs}\footnote{\url{https://github.com/spacetelescope/tweakwcs}} and then the individual images were aligned to the mosaic image. The calibrated and aligned files were then passed through the Stage 3 pipeline to create drizzled images in each bands. During the resampling step, we used a fixed pixel scale of 0.031\arcsec\, for SW images and 0.0315\arcsec\, for LW images with {\tt pixfrac=0.8}. To improve the astrometry of the mosaic images,  we further aligned all drizzled images for each quasar field to GAIA DR3. 
{The aligned mosaic images have a  precise astrometry and the comparison of NIRCam positions with Gaia catalog typically reveals small offsets of $\sim$50 mas after considering proper motions.}
Finally, we derived and subtract a background for each mosaic using {\tt photutils}\footnote{\url{https://photutils.readthedocs.io}}.

Unlike \citet{Wang23}, we use {\tt SourcExtractor++} \citep{Bertin20} instead of {\tt SExtractor} \citep{SExtractor} for source detection. Detailed descriptions of catalog construction can be found in \cite{Champagne24a,Champagne24b}, but we also briefly summarize it in the following. First, we performed PSF matching on F115W and F200W images to match the resolution of F356W images using empirical PSF models. The empirical PSF models for each images were created by stacking GAIA stars (if available) and bright point sources detected by {\tt daophot} from the {\tt photutils} packages in each image. After matching PSF, we used the {\tt SourcExtractor++} dual-image mode with the respective weight maps for each filter with F356W as the detection band to extract photometry. We set {\tt DETECT\_THRESH=3.0}, Kron parameters {\tt k, Rmin = 1.2, 1.7}, {\tt DETECT\_MINAREA=10} pixels, and {\tt PHOT\_APERTURES= 12 } pixels. 
{To derive an aperture correction for each source, we ran {\tt SourcExtractor++} twice, once with custom Kron settings optimized for small sources ({\tt k, Rmin = 1.2, 1.5}) and then the default Kron settings ({\tt k, Rmin=2.5, 3.5}). The aperture correction is the ratio between these two Kron fluxes measured in the F356W filter and applied to the photometry in every filter. The final photometry was measured as the forced aperture photometry multiplied by the aperture correction ({\tt Kron\_custom/Kron\_default}) to account for flux outside of the circular aperture.}
The Galactic extinctions in each filter were corrected on the basis of the dust reddening map from \cite{schlegel98} and the dust extinction law derived from \cite{Schlafly11}. The signal-to-noise ratios are measured in a $0\farcs32$ arsec circular aperture, as listed in Table \ref{tbl:depth}. To determine the photometric uncertainties and limiting magnitudes in each ASPIRE fields, we placed 1000 empty apertures across the images after masking out real sources using the segmentation map, and measured the median absolute deviation of flux in the fixed aperture size. The entire workflow, including the NIRCam/WFSS data reduction and emission line searching algorithms described in the following sections, is integrated into a dedicated Python package, \texttt{unfold\_jwst} (Wang et al., in prep.). 
{The observing dates and the achieved $5\sigma$ limiting magnitudes, measured in $0\farcs32$ apertures for each of the ASPIRE quasar fields, are listed in Table~ \ref{tbl:depth}.}

\subsection{NIRCam WFSS data reduction}\label{sec:wfss_reduction}
Similar to the reduction of imaging data, the WFSS data were reduced using the combination of \texttt{CALWEBB} (version 1.10.2) and some custom scripts in \texttt{unfold\_jwst}. We use calibration reference files (\texttt{jwst\_1080.pmap}) from version 11.16.21 of the standard Calibration Reference Data System (CRDS). We refer the reader to \citet{Wang23} and \citet{Yang23} for a more detailed description of the process.
Briefly, we used the \texttt{CALWEBB} stage 1 pipeline to calibrate the detector-level signals and the ramp fitting for individual WFSS exposures. We then subtracted the $1/f$ noise pattern along each detector column. The $1/f$ noise-subtracted images were flat fielded and assigned {\tt WCS} using routines in the \texttt{CALWEBB} pipeline. We also subtracted master median background models from the processed images. After that, we measured astrometric offsets between each of the SW images obtained simultaneously with WFSS exposures and the fully calibrated F356W mosaic. The offsets were applied to the tracing and dispersion models \citep{Sun22a, Sun22b} when extracting the spectra. 
We extracted both 2D and 1D spectra for all sources detected in the F356W imaging. The 2D spectrum of each source is extracted from each individual exposure, and the individual 2D spectra are then coadded into a 2D spectrum after resampling them to common wavelength (9.8 $\rm \AA~pixel^{-1}$) and spatial grids following the histogram2D technique in the \texttt{PypeIt} software \citep{Pypeit,Pypeit_ascl}.
We then extracted 1D spectra from the coadded 2D spectra using both the optimal \citep{Horne86} and the boxcar extraction algorithms. We used an aperture diameter of 7 pixels ($0\farcs411$) for the boxcar spectral extraction. 
In addition to extracting a 1D spectrum from the coadded 2D spectrum for each source, we optimally extracted 1D spectra from all individual 2D spectra. The extracted 1D spectra from individual exposures were then combined with inverse variance weighting and outlier rejection to a stacked 1D spectrum for each source. The main purpose for producing such stacked spectra is that such a method provides better removal of hot pixels or other outliers for sources with strong continuum (i.e., allow for a profile fitting for optimal extraction from individual exposures). 

\begin{figure}
\centering
\includegraphics[width=0.49\textwidth]{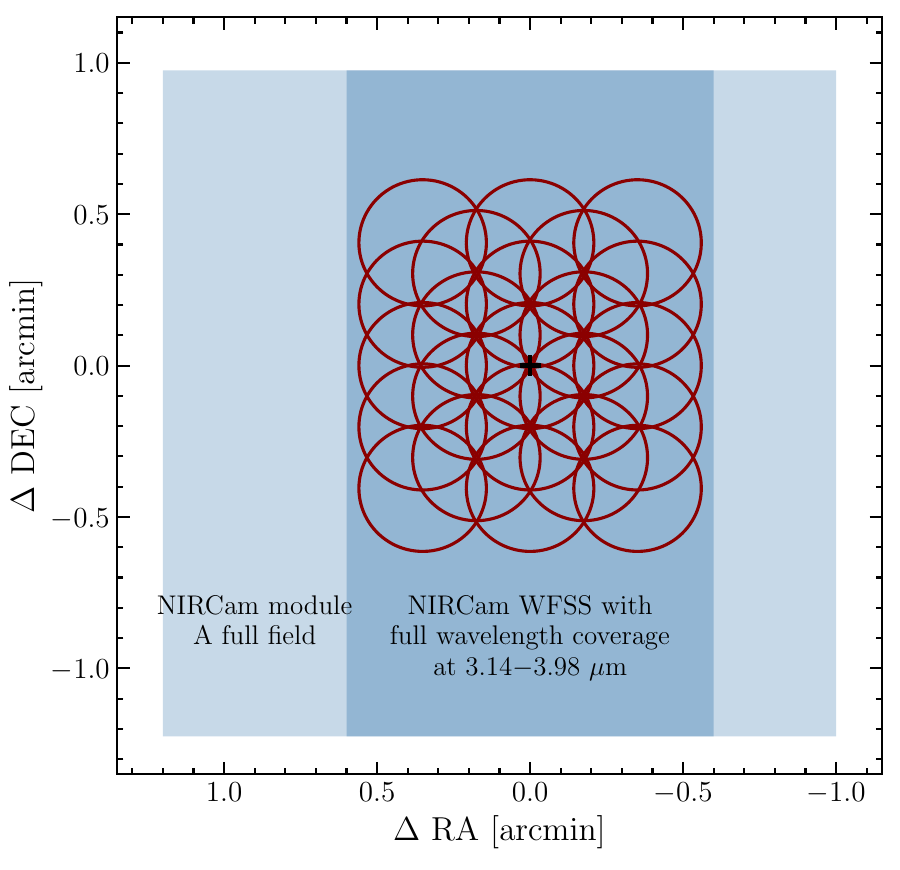}
\caption{{\bf ALMA mosaic pattern.}
We used a 23-pointings ALMA mosaic observation to cover a $\sim1.3$ arcmin$^2$ sky area (highlighted by red circles) for each quasar field. This region falls into the full wavelength coverage footprint of the NIRCam module A slitless spectroscopic observation (blue shaded region). Note that the actual NIRCam observations have different position angles for different fields, while all ALMA mosaic were observed at position angle zero. The black cross marks the position of the central quasar. 
\label{fig:footprint}}
\end{figure}

\subsection{ALMA observations and data reduction}\label{sec:alma_obs}
To discover dusty galaxies in quasar environments beyond the one-megaparsec scale, we designed an ALMA mosaic program (2022.1.01077.L, PI: F. Wang), which consists of 23 Nyquist-sampled pointings in Band 6 (Figure \ref{fig:footprint}). This mosaic covers $\sim1.4$ arcmin$^2$ (or $\sim9$ cMpc$^2$ at $z\gtrsim6.5$) projected sky area around each quasar. In addition, the ALMA footprint falls into the full wavelength coverage region of NIRCam module A, which ensures that we could potentially determine the redshifts for dusty star-forming galaxies by combing ALMA observations with JWST NIRCam/WFSS observations (Figure \ref{fig:footprint}). 

Our configuration is motivated by searching for [\ion{C}{2}] line emitters at the quasar redshifts. To achieve this goal, we tune two spectral windows (SPW, 0 and 1) centered at the expected frequency of [\ion{C}{2}] (LSB or lower side band). This enables us to search for [\ion{C}{2}] line emitters physically associated ($\Delta v \lesssim \pm2000~{\rm km~s^{-1}}$) with the central quasars. We tune the other two SPWs (2 and 3) centered at about 15 GHz away from the expected [\ion{C}{2}] line (USB or upper side band) for observing continuum emission and searching for [\ion{C}{2}] or CO line emitters at different redshifts. We choose a spectral line setup, with 8 channel averaging, which leads to a velocity resolution of $\sim 5~{\rm km~s^{-1}}$. 

The ALMA observation is designed to detect $z>6.5$ galaxies with [\ion{C}{2}]-based star formation rate (SFR) of $\rm SFR\gtrsim30~{\rm M_\odot~yr^{-1}}$. This leads to $\sim4$ hr total time for each ASPIRE quasar field and a total time of $\sim$100 hours for the entire ALMA program. To balance maximizing the sensitivity and avoiding confusion when matching ALMA data with JWST, we chose to use C-2/3 configuration which delivers a resolution of 0\farcs5--1\farcs0. The ALMA observations were conducted from October 14, 2022 to February 14, 2023. The typical angular resolution of ALMA observation is $\sim$\,0\farcs65. The total survey area of the entire ALMA program is 34.9\,arcmin$^2$ at above a primary beam response limit of 0.25. 

All the ALMA data were calibrated and processed with the standard {\tt CASA v6.4.1.12} pipeline \citep{CASA}. Before making the continuum map, we visually identified and flagged the spectral channels affected by telluric absorption using the {\tt plotms} command. The continuum map was imaged with the {\tt tclean} command with a cell size of 0\farcs12. We used the automatic multi-threshold masking algorithm during the imaging process. To accommodate both compact sources and extended sources, we produced two sets of continuum maps. The first set of continuum maps was produced in the native ALMA resolution with a Briggs weight parameter of {\tt robust=0.5} and the second set of continuum maps was produced with {\tt robust=2.0} and tapered with a Gaussian kernel of FWHM\,=\,1\farcs0. The resulting median synthesized beam (FWHM) is about $0\farcs65$ and $1\farcs34$ for native and tapered continuum images, respectively. The continuum RMS noises of our ALMA mosaics are 0.031$\pm$0.004\,mJy\,beam$^{-1}$ and 0.034$\pm$0.004\,mJy\,beam$^{-1}$ at native and tapered resolution, respectively. Before creating data cubes, we subtract the continuum with {\tt fitorder=1}. The data cubes were created with a channel width of 30 km s$^{-1}$ using {\tt robust=2.0}, {\tt clark} deconvolver, $2\sigma$ threshold and a mask with a radius of 1\farcs0 centered on the central quasars. A more detailed ALMA reduction process can be found in \cite{Sun25} and Decarli et al. in prep. 

\begin{figure}
\centering
\includegraphics[trim=460 10 460 10, clip, width=0.5\textwidth]{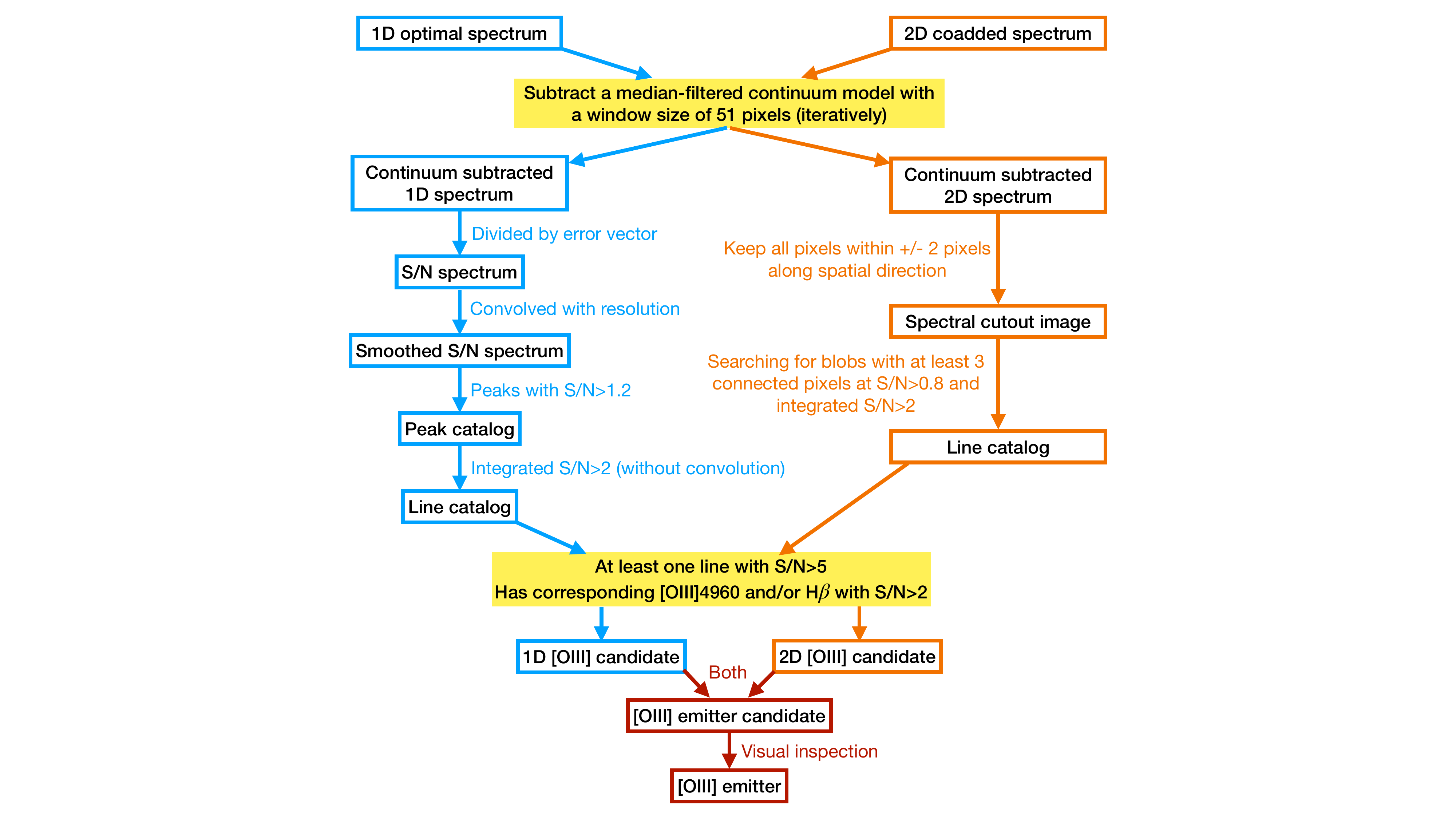}
\caption{{\bf [\ion{O}{3}] emitter searching strategy.}
 The left panel shows the [\ion{O}{3}] emitter detection algorithm based on 1D spectra, while the right panel illustrates the algorithm based on 2D spectra. We visually inspected all [\ion{O}{3}] emitter candidates identified by both algorithms and compiled the final [\ion{O}{3}] emitter catalog.
 \label{fig:algorithm}}
\end{figure}

\section{Line emitting galaxies}\label{sec:emitters}
\subsection{[\ion{O}{3}] emitter finding algorithm}\label{subsec:o3}
The NIRCam/WFSS observations allow us to search for line emitters in an unbiased way without any imaging preselection. In this work, we extract spectra for all sources detected in the F356W band and then identify [\ion{O}{3}] emitters based on the extracted spectra directly. 
{Based on the emission-line searching algorithm developed by \cite{Wang23}, we further improved the efficiency and purity of the emission-line detections and performed a semiautomatic search using the continuum-subtracted 2D and 1D spectra. For the 1D case, we generated a median-filtered continuum model with a window size of 51 pixels (corresponding to $\sim0.05~\mu$m), which was then subtracted from the optimally extracted spectrum. \cite{Wang23} searched for peaks directly in the continuum-subtracted 1D spectra, which introduced a large number of hot pixels. To address this issue, we instead searched for $\mathrm{S/N} > 1.2$ peaks in a smoothed S/N spectrum rather than in the raw 1D spectrum. The smoothed S/N spectrum was obtained by convolving the continuum-subtracted S/N spectrum with the spectral resolution ($\sim 2.3$ pixels) of the WFSS observations. We found that peaks identified in the smoothed spectra could still be contaminated by hot pixels or other artifacts (e.g., residuals caused by imperfect contamination-model subtraction or background subtraction).
} 
To remove such artifacts, we only treat peaks with a total S/N of $\rm S/N>2$ as emission line candidates, where the total S/N was measured by integrating $\pm2$ original pixels (i.e., the S/N spectra without convolving with the spectral resolution) from the center of the peak. 
To identify if a certain object is a [\ion{O}{3}] emitter candidate, we first assume that all identified lines with $\rm S/N>5$ (if exists) are the [\ion{O}{3}] $\lambda5008$ line and then ask if a corresponding [\ion{O}{3}] $\lambda4960$ or H$\beta$ line also exists (i.e., with $\rm S/N>2$). If [\ion{O}{3}] $\lambda4960$ and/or H$\beta$ line exists, we then treat this object as a [\ion{O}{3}] emitter candidate. 

\begin{figure}
\centering
\includegraphics[trim=0 0 0 0, clip, width=0.5\textwidth]{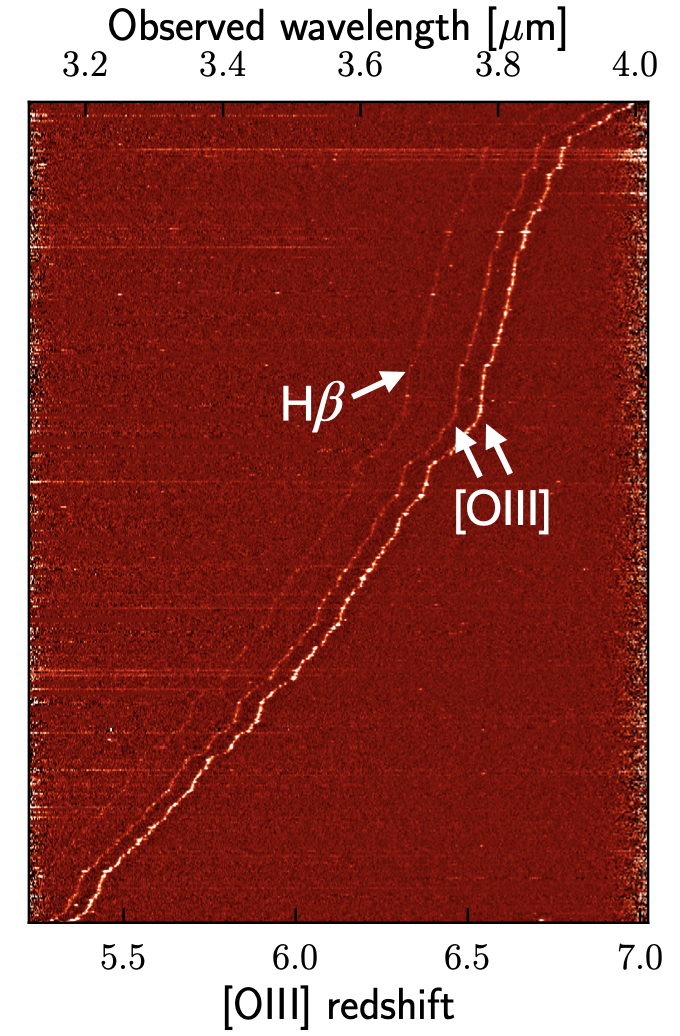}
\caption{{\bf Spectra of 487 [\ion{O}{3}] emitters.}
 Each row displays the spectrum of a single [\ion{O}{3}] emitter. The three bright spots from left to right correspond to H$\beta$, [\ion{O}{3}] $\rm\lambda4960,\text{\AA}$, and [\ion{O}{3}] $\rm\lambda5008,\text{\AA}$, respectively. In some cases, additional scattered emission lines from contaminating sources are also visible.
\label{fig:spec2d}}
\end{figure}

\begin{figure}
\centering
\includegraphics[width=0.5\textwidth]{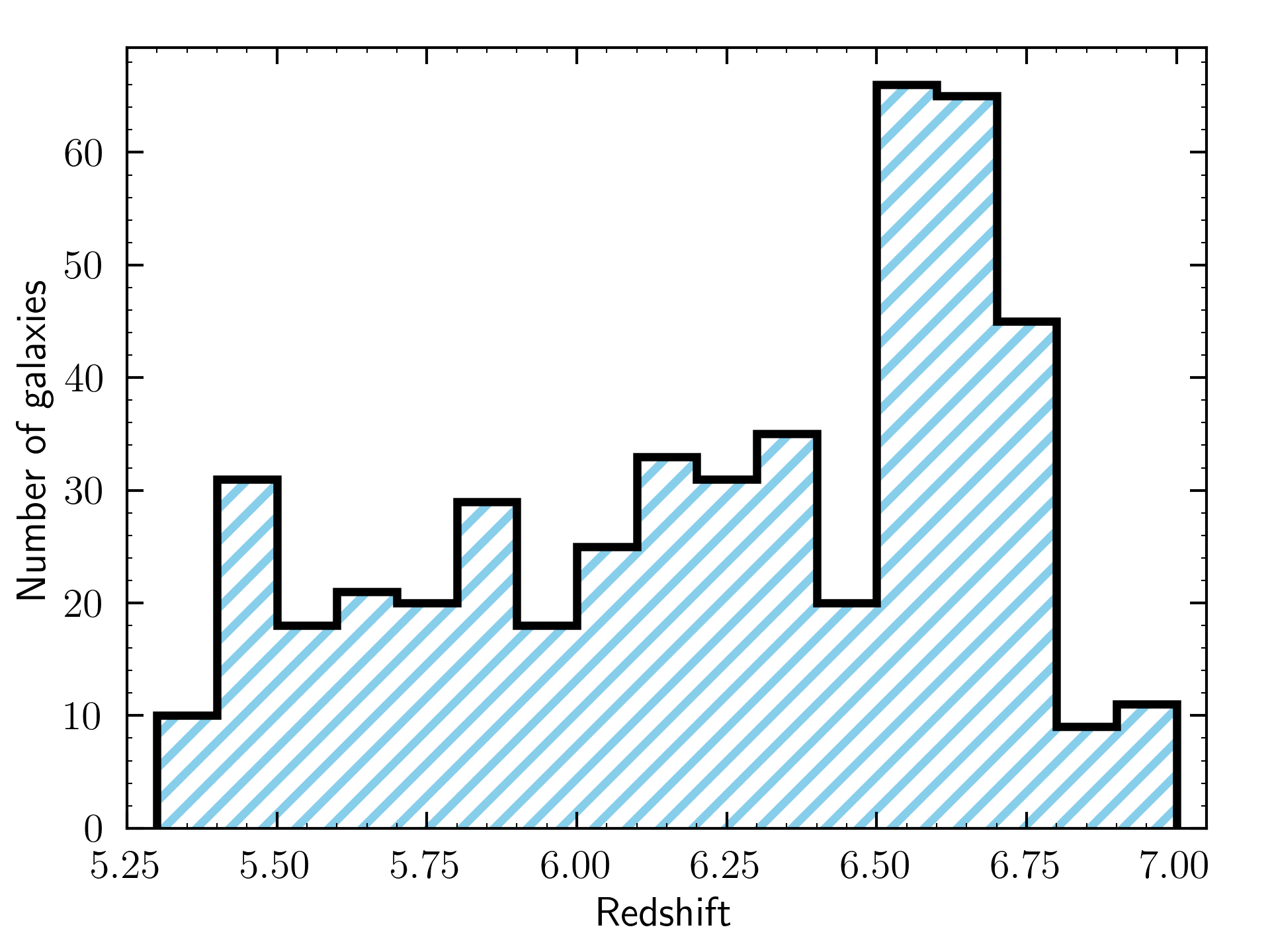}
\caption{{\bf The redshift distribution of [\ion{O}{3}] emitters.}
 ASPIRE identified 487 [\ion{O}{3}] emitters at $5.3 \lesssim z \lesssim 7$. A significant excess of galaxies at $6.5 < z < 6.8$, coinciding with the redshifts of ASPIRE quasars, is clearly shown in this figure, indicating that ASPIRE quasars typically trace galaxy overdensities.
 \label{fig:redshift}}
\end{figure}

We also search for [\ion{O}{3}] emitter candidates from the continuum subtracted 2D spectra independently. We first search for bright blobs from coadded 2D spectra using {\tt Photutils} \citep{photutils}. For each object, we identified bright blobs on a 2D spectral cutout image with a diameter of 5 pixels perpendicular to the dispersion direction. We treat all blobs that have at least three connected pixels with $\rm S/N>0.8$ and the integrated flux with $>2\sigma$ significance as emission line candidates. We then search for potential [\ion{O}{3}] emitter candidates from the 2D emission line catalogs using the algorithm developed above (i.e., [\ion{O}{3}] $\lambda5008$ with $\rm S/N>5$ and [\ion{O}{3}] $\lambda4960$ or H$\beta$ with $\rm S/N>2$).  

\begin{figure*}
\centering
\includegraphics[width=0.99\textwidth]{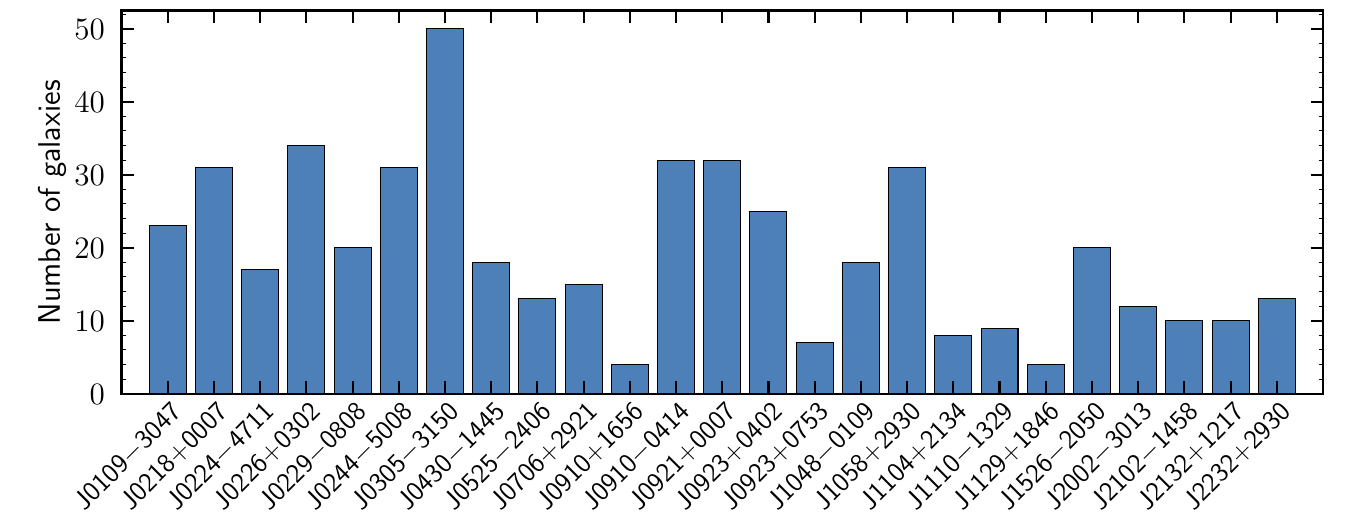}
\caption{{\bf Number of [\ion{O}{3}] emitters identified in each ASPIRE quasar field.}
 Each quasar sightline covers a survey volume of approximately $\sim44,000~\mathrm{cMpc}^3$, spanning from $z \sim 5.3$ to $z \sim 7$. The number of galaxies varies significantly across fields, indicating substantial cosmic variance in single-pointing JWST observations.
 \label{fig:ngal_all}}
\end{figure*}

To ensure the robustness of the [\ion{O}{3}] emitter candidate identification, we only kept objects that were identified as [\ion{O}{3}] emitter candidates from both the 1D and 2D algorithms. This requirement also significantly reduced the required visual inspection efforts. 
To reduce the bias of visual inspection, we required that each object be inspected by at least three different team members (E.B., J.B.C., X.J., K.K., R.A.M., F.W., Y.W., J.Y.). During the first round of our visual inspection, each object was classified to be one of the following four categories:  1) {\tt score=3} means a target is definitely an [\ion{O}{3}] emitter, 2) {\tt score=2} means a target could be an [\ion{O}{3}] emitter but the corresponding [\ion{O}{3}] $\lambda$4960 line is not detected and the corresponding H$\beta$ is very faint, 3) {\tt score=1} means only one emission line was detected for the source, 4) {\tt score=0} means a target is definitely not an [\ion{O}{3}] emitter. After the first round inspection, we classify sources with at least two {\tt score=3} as high confidence [\ion{O}{3}] emitters, sources with at least one {\tt score=3} or at least two {\tt score=2} as [\ion{O}{3}] emitter candidates, and other sources as unlikely to be [\ion{O}{3}] emitters. 
{ Furthermore, we asked the inspectors to re-evaluate whether the first two classes (i.e., high-confidence [\ion{O}{3}] emitters versus [\ion{O}{3}] emitter candidates) should be interchanged, taking into account comments from other inspectors regarding the confidence of the line detections and potential line-confusion issues. Emitter candidates for which concerns were raised by more than one inspector about possible line confusion and/or significant contamination were reassigned to the lower-confidence [\ion{O}{3}] emitter-candidate category.}
To ensure a high purity of the catalog, we decided to use only high-confidence [\ion{O}{3}] emitters in this work. In future work, we will combine photometric redshifts derived from both JWST and ground-based imaging with our current catalog of [\ion{O}{3}] emitter candidates to construct a more complete [\ion{O}{3}] emitter sample at the faint end. The [\ion{O}{3}] emitter search strategy is summarized in Figure~\ref{fig:algorithm}. We found that this strategy can recover all [\ion{O}{3}] emitters except for the faintest one (ASPIRE-J0305M31-O3-023) in \cite{Wang23} and identified 10 additional new [\ion{O}{3}] emitters in this field. 
{By re-inspecting the spectra from \cite{Wang23}, we found that these ten newly identified [\ion{O}{3}] emitters exhibit relatively weak line detections and were missed in the earlier work primarily due to the slightly lower data sensitivity caused by different version of data reduction pipeline and/or the differences in the peak-detection algorithm used by \cite{Wang23}. }
This indicates that our new line emitter searching method is more effective.

\subsection{A catalog of 487 [\ion{O}{3}] emitters}\label{subsec:o3e}
Using the strategy discussed in the previous section, we successfully identified 487 high-confidence [\ion{O}{3}] emitters at $5.2\lesssim z\lesssim 7$. We show the spectra of all 487 [\ion{O}{3}] emitters in Figure~\ref{fig:spec2d}. A detailed characterization and the luminosity function of these spectroscopically confirmed [\ion{O}{3}] emitters will be presented in a companion paper (Champagne et al., in prep). In the following sections, we focus our discussions on the quasar environment.

Figure \ref{fig:redshift} shows the redshift distribution of these [\ion{O}{3}] emitters. From this plot, we can clearly see a number excess of [\ion{O}{3}] emitters at $6.5<z<6.8$, the redshift range of ASPIRE quasars. This excess is also evident in Figure~\ref{fig:spec2d}, where a noticeable steepening in the distribution of [\ion{O}{3}] emission lines appears at $6.5 < z < 6.8$. There are in total 122 [\ion{O}{3}] emitters with a line-of-sight velocity differencing of $\Delta v_{\rm los}<1000~{\rm km~s^{-1}}$ relative to the central quasars. {The $\Delta v_{\rm los}<1000~{\rm km~s^{-1}}$ criterion is widely used in the literature, which corresponds to $\sim9.5$ cMpc at $z\sim6.6$, comparable to the largest projected distance probed by the NIRCam field of view.} 
After accounting for variations in survey volume and sensitivity across different locations and wavelengths (or redshifts), we find that the number density of galaxies at the quasar redshifts ($|\Delta v_{\rm los}| < 1000~{\rm km~s^{-1}}$, $ n_{{\rm gal}, z_{\rm quasar}}= 10^{-2.33^{+0.04}_{-0.04}}~{\rm Mpc^{-3}}$) is approximately 9.4 times higher than that at other redshifts ($ n_{\rm gal}= 10^{-3.31^{+0.02}_{-0.02}}~{\rm Mpc^{-3}}$). Such a significant overdensity strongly indicates that the ASPIRE quasars are excellent tracers of galaxy overdensities and may serve as reliable probes of massive dark matter halos in the early Universe.
We note that the average galaxy number density at non-quasar redshifts is approximately twice as low as that measured in the ASPIRE J0305--3150 field by \citet{Wang23}. To investigate this discrepancy, we present in Figure~\ref{fig:ngal_all} the number of [\ion{O}{3}] emitters identified in each ASPIRE field. A striking result is the large field-to-field variation in the number of detected galaxies, despite each ASPIRE field probing a survey volume of $\sim44,000~{\rm cMpc^3}$ over $5.2 \lesssim z \lesssim 7$. This finding underscores the strong influence of cosmic variance in pencil-beam surveys, especially those relying on single JWST pointings or only a few sightlines. Such large field-to-field variations may also explain the inconclusive results of quasar environment studies over the past two decades (see \S\ref{sec:environment} for further discussion). Because the uncertainty in clustering measurements scales approximately with the intrinsic variance of the density field divided by the square root of the number of independent fields, ASPIRE's 25 sightlines (compared to only four in \citealt{Eilers24}) provide a significant improvement in constraining quasar environments and clustering properties. In \S\ref{sec:clustering}, we present a clustering analysis based on the [\ion{O}{3}] emitters identified across these 25 ASPIRE quasar fields.

\subsection{The discovery of 17 [\ion{C}{2}] emitters from ALMA data}
\setlength{\tabcolsep}{2pt}
\begin{deluxetable}{llllrrr}
\vspace{15pt}
\tablehead{\colhead{Name} & \colhead{RA} & \colhead{DEC} & \colhead{$z$} & \colhead{S/N} & \colhead{$v_{\rm los}$} & \colhead{Dist}}
\startdata
& & & & & km/s & kpc\\
\hline
J0226$+$0302.C2 & 02:26:00.92 & $+$03:03:17.03 & 6.5430 & 7.1 & 98.6 & 126.5 \\
J0226$+$0302.C3 & 02:26:00.93 & $+$03:03:16.74 & 6.5400 & 6.3 & $-$20.4 & 124.7 \\
J0229$-$0808.C1 & 02:29:37.55 & $-$08:08:36.78 & 6.7004 & 6.6 & $-$951.8 & 201.9 \\
J0229$-$0808.C2 & 02:29:35.32 & $-$08:08:22.70 & 6.7325 & 6.2 & 293.2 & 6.4 \\
J0244$-$5008.C1 & 02:44:00.16 & $-$50:08:39.30 & 6.7307 & 8.7 & 259.8 & 90.7 \\
J0244$-$5008.C2 & 02:43:58.98 & $-$50:08:27.71 & 6.3142 & 7.0 & $-$15906.6 & 178.1 \\
J0305$-$3150.C1 & 03:05:16.39 & $-$31:50:55.14 & 6.6077 & 8.2 & $-$243.5 & 37.4 \\
J0430$-$1445.C1 & 04:30:42.52 & $-$14:45:40.97 & 6.6843 & 9.6 & $-$1162.9 & 90.4 \\
J0525$-$2406.C1 & 05:25:59.75 & $-$24:06:23.45 & 6.5373 & 19.4 & $-$94.4 & 6.3 \\
J0706$+$2921.C1 & 07:06:26.50 & $+$29:21:14.44 & 6.5704 & 6.6 & $-$1314.8 & 50.4 \\
J0910$-$0414.C1 & 09:10:53.60 & $-$04:14:09.13 & 6.6236 & 22.5 & $-$500.4 & 77.8 \\
J0910$-$0414.C2 & 09:10:54.54 & $-$04:13:55.52 & 6.6292 & 15.7 & $-$276.9 & 62.4 \\
J0923$+$0402.C1 & 09:23:44.83 & $+$04:02:41.35 & 6.6340 & 6.1 & 37.9 & 202.5 \\
J1110$-$1329.C1 & 11:10:34.02 & $-$13:29:46.25 & 6.5081 & 6.1 & $-$249.5 & 6.1 \\
J1526$-$2050.C1 & 15:26:37.87 & $-$20:50:02.33 & 6.5923 & 11.5 & 214.9 & 9.5 \\
J1526$-$2050.C2 & 15:26:37.20 & $-$20:50:31.63 & 6.5954 & 6.0 & 334.8 & 178.5 \\
J2232$+$2930.C1 & 22:32:56.22 & $+$29:31:05.30 & 6.6575 & 6.2 & $-$333.8 & 198.1
\enddata
\caption{Identified high confidence [\ion{C}{2}] emitter candidates.}
\label{tbl:c2}
\end{deluxetable}

To understand the distribution of dusty galaxies in the quasar vicinity, we searched for galaxies with [\ion{C}{2}] emission lines from the ASPIRE ALMA mosaic observations. The detailed description of the [\ion{C}{2}] emission line searching will be presented in Decarli et al. in prep. Briefly, the [\ion{C}{2}] line emitters were identified using the \texttt{INTERFEROPY} software \citep{Interferopy}, which is a \texttt{Python} implementation of \texttt{FINDCLUMPs} \citep{Walter16}. \texttt{INTERFEROPY} runs SExtractor \citep{SExtractor} on averaged channels of continuum subtracted cubes with a range of kernel widths and identifies objects with S/N$>$3. Each line emitter candidate was assigned with a \textit{fidelity} by comparing the candidate's S/N with the S/N distribution of positive and negative candidates for a given kernel width. In this work, we consider only line emitter candidates with S/N$>6$, a threshold at which the fidelity reaches $\sim$100\%, ensuring a highly pure galaxy sample suitable for clustering analysis. 

In total, we identify 17 positive line emitter candidates with S/N$>6$, 16 of which lie within the line search bandwidth (LSB) corresponding to the expected [\ion{C}{2}] wavelength at the quasar redshifts. This strongly suggests that these sources are highly likely [\ion{C}{2}]-emitting galaxies associated with the quasars. The basic properties of these [\ion{C}{2}]-emitting galaxies are summarized in Table \ref{tbl:c2}. For a more detailed discussion of the individual line emitters, we refer the reader to Decarli et al.\ (in prep.). In this work, we focus on the statistical clustering analysis of these 15 ALMA [\ion{C}{2}] emitter candidates and the [\ion{O}{3}] emitters identified in \S \ref{subsec:o3e} to probe the large-scale environments of the quasar fields.

\setlength{\tabcolsep}{5pt}
\begin{deluxetable}{ccccc}
\tablehead{\colhead{$R_{\rm min}$} & \colhead{$R_{\rm max}$} & \colhead{$\left <QG \right >$} & \colhead{$\left <QR \right >$} & \colhead{$\chi(R_{\rm{min}}, R_{\rm{max}})$}}
\startdata
${\rm cMpc}\,h^{-1}$ & ${\rm cMpc}\,h^{-1}$ & & &\\
\hline
0.060 & 0.106 & 3 & 0.0082 & $364.5^{+355.5}_{-198.9}$ \\
0.106 & 0.186 & 3 & 0.0367 & $80.8^{+79.6}_{-44.5}$ \\
0.186 & 0.329 & 6 & 0.1344 & $43.6^{+26.7}_{-17.7}$ \\
0.329 & 0.580 & 11 & 0.4359 & $24.2^{+10.1}_{-7.5}$ \\
0.580 & 1.022 & 34 & 1.3353 & $24.5^{+5.2}_{-4.3}$ \\
1.022 & 1.802 & 36 & 2.7924 & $11.9^{+2.5}_{-2.1}$ \\
1.802 & 3.176 & 21 & 2.7692 & $6.6^{+2.0}_{-1.6}$ \\
3.176 & 5.600 & 4 & 1.2167 & $2.3^{+2.6}_{-1.6}$ \\
\hline
0.025 & 0.079 & 4 & 0.0035526 & $1124.9^{+890.2}_{-538.8}$\\
0.079 & 0.251 & 1 & 0.0355269 & $27.15^{+64.73}_{-23.29}$\\
0.251 & 0.794 & 7 & 0.3552692 & $18.70^{+10.61}_{-7.27}$\\
0.794 & 2.512 & 4 & 0.4819394 & $7.30^{+6.56}_{-3.97}$
\enddata
\caption{Quasar-galaxy cross-correlation function measurements. The upper portion corresponds to results for [\ion{O}{3}] emitters, while the lower portion corresponds to those for [\ion{C}{2}] emitters.}
\label{tbl:ccf}
\end{deluxetable}

\section{Clustering analyses}\label{sec:clustering}
\subsection{Quasar-galaxy cross-correlation function}
To determine the clustering of galaxies around ASPIRE quasars, we follow \cite{hennawi06b} and measure a volume-averaged projected cross-correlation function between quasars and line emitters (both [\ion{O}{3}] and [\ion{C}{2}] emitters) defined by
\begin{equation}
    \chi(R_{\rm{min}}, R_{\rm{max}}) = \frac{\int \xi_{QG}(R, Z) dV_{\rm{eff}}}{V_{\rm{eff}}} \ ,
\label{eq:ccf}    
\end{equation}
where $\xi_{QG}(R, Z)$ is the real-space quasar-line emitter two-point correlation function and $V_{\rm{eff}}$ is the effective volume of the cylindrical shell between $R_{\rm{min}}$ and $R_{\rm{max}}$ and extending to a cylinder height $\pm Z$,
\begin{equation}
Z = \frac{c}{H(z)}\delta z \ ,
\end{equation}
where $H(z)$ is the Hubble constant at redshift $z$. 
The volume averaged cross-correlation $\chi(R_{\rm{min}}, R_{\rm{max}})$ is calculated in radial bins with bin edges $R_{\rm{min}}, R_{\rm{max}}$ via 
\begin{equation}
    \chi(R_{\rm{min}}, R_{\rm{max}}) = \frac{\langle QG\rangle}{\langle QR \rangle} -1 \ ,
\end{equation}
where $\langle QG\rangle$ is the number of quasar-emitter pairs in the enclosed cylindrical volume, which is directly measured by counting the quasar-line emitter pairs found in ASPIRE data, and $\langle QR \rangle$ is the expected number of quasar-line emitter pairs in the same bin if they were randomly distributed around the quasar, with the background number density. 
The $\langle QR \rangle$ is computed from 
\begin{equation}
   \langle QR \rangle = n_{\rm G} (z) V_{\rm eff},
\end{equation}
where $n_{\rm G} (z)$ is the mean number density of line emitters at redshift $z$ within our survey. 

Since the ALMA mosaic observations provide uniform survey depth in both the lower and upper sidebands (LSB and USB), corresponding to galaxies at the quasar redshifts and at slightly lower redshifts, respectively, we can directly estimate the field galaxy number density, $n_{\rm G}(z)$, from the USB data. After accounting for the survey volume of each ASPIRE quasar field and assuming that all line emitters with S/N$>$6 are [\ion{C}{2}] emitters with a redshift-independent number density across the narrow redshift range of interest, we measure $n_{\rm G}(z) = 10^{-3.98^{+0.76}_{-0.52}}~{\rm cMpc^{-3}}$ using galaxies within USB. Given the limited number of [\ion{C}{2}] emitter candidates, we use all galaxies in LSB to measure $\langle QG \rangle$, corresponding to a cylinder height of $Z\sim 14~{\rm cMpc}\,h^{-1}$ or line-of-sight velocity $\Delta v_{\rm los} \sim \pm2100 \, \text{km s}^{-1}$.
The $\chi(R_{\rm{min}}, R_{\rm{max}})$ was measured in four bins and the results are presented in Table \ref{tbl:ccf}.

The clustering analysis for [\ion{O}{3}] emitters is more complex than that for [\ion{C}{2}] due to the strong dependence of the NIRCam/WFSS line-flux limit on both wavelength (or redshift) and spatial position. To fully account for these spatial and spectral variations, we performed mock source injection experiments across all ASPIRE quasar fields to determine the corresponding flux limits as functions of both wavelength and position. Firstly, we generated a grid of mock sources at $5.3<z<7$ with a fixed line flux ($f = 10^{-17}~\rm erg\,s^{-1}\,cm^{-2}$, five times brighter than our flux limit determined by \cite{Wang23}) and applied the same line-searching algorithm described in \S \ref{subsec:o3}. 
The flux limit at each wavelength and position was then derived from the S/N of the detected emission lines. Using these line flux limit maps, we then constructed an \texttt{unfold\_jwst} detected random source catalog and a corresponding selection function by injecting five million ($5\times10^{6}$) mock sources into the full F356W imaging footprint of each quasar field and identifying which mock sources met the detection threshold of the emitter search. The flux distribution of the injected sources was matched to the [\ion{O}{3}] luminosity function from \cite{Matthee22}, assuming no redshift evolution. 
We then used the \texttt{Corrfunc} python package \citep{Sinha20} to count quasar-galaxy pairs, $\langle QG \rangle$, and quasar-mock galaxy pairs, $\langle QR \rangle$, within cylindrical bins defined by $(R_{\rm min}, R_{\rm max})$ and a cylinder height of $Z = 7~{\rm cMpc}\,h^{-1}$, corresponding to a line-of-sight velocity of $\Delta v_{\rm los} \lesssim 1000~{\rm km~s^{-1}}$ at the ASPIRE quasar redshifts. The $\langle QR \rangle$ values were normalized using the expected number of [\ion{O}{3}] emitters in blank fields after applying the selection function. A comprehensive description of this methodology and the measurement of [\ion{O}{3}] emitter auto-correlation function will be presented in a companion paper by \cite{Huang26}. The resulting $\chi(R_{\rm min}, R_{\rm max})$ measurements for the [\ion{O}{3}] emitters are listed in Table~\ref{tbl:ccf}. 
Note that the uncertainties of all $\chi(R_{\rm min}, R_{\rm max})$ measurements in Table~\ref{tbl:ccf} include only Poisson noise. In \cite{Huang26}, we will further incorporate the effects of cosmic variance into the error budget using a covariance matrix constructed from mock realizations of cosmological simulations.

\begin{figure}
\centering
\includegraphics[width=0.49\textwidth]{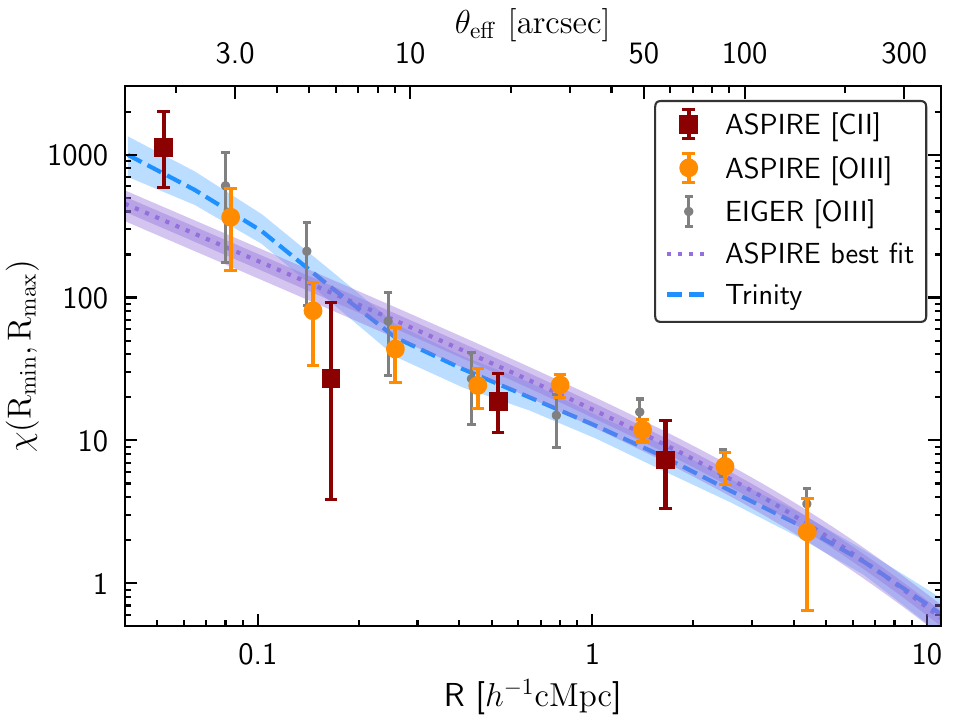}
\caption{{\bf  Quasar-galaxy cross-correlation function.}
The orange points represent the measured quasar-galaxy cross-correlation function based on [\ion{O}{3}] emitters while the dark red points denote the cross-correlation function from [\ion{C}{2}] emitters. 
The purple dashed line and dark (shallow) shaded region show the power-law fit when fixing the slope $\gamma_{\rm{QG}} = 2.0$ and its associated $1\sigma$ ($2\sigma$) posterior range. The blue dashed line show the predictions based on \texttt{Trinity} model. The gray dots are quasar-galaxy cross-correlation function measured by the EIGER program using only four quasars at $z\sim6$ \citep{Eilers24}. 
\label{fig:ccf}}
\end{figure}

\begin{figure}
\centering
\includegraphics[width=0.49\textwidth]{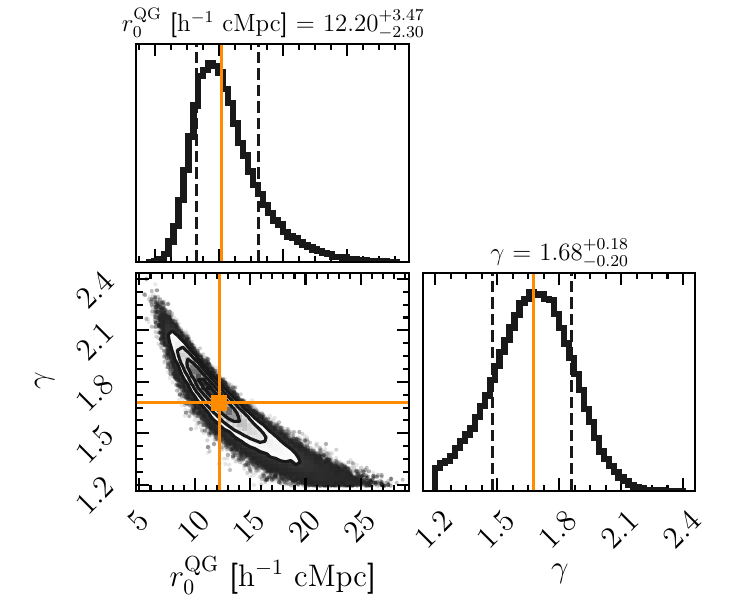}
\caption{{\bf  The corner plot of the joint fitting for $\gamma_{\rm{QG}}$ and $r_0^{\rm{QG}}$.}
We derive $\gamma_{\rm{QG}} = 1.68^{+0.18}_{-0.20}$ and $r_0^{\rm{QG}} = 12.20^{+3.47}_{-2.30}~h^{-1}~{\rm cMpc}$ without fixing $r_0^{\rm{QG}}$, however, there is a strong degeneracy between  $\gamma_{\rm{QG}}$ and $r_0^{\rm{QG}}$.
\label{fig:corner}}
\end{figure}

We show the measured cross-correlation function, $\chi(R_{\rm{min}}, R_{\rm{max}})$, in figure \ref{fig:ccf}. With Eq. \ref{eq:ccf}, we can fit $\chi(R_{\rm{min}}, R_{\rm{max}})$ in order to constrain the real-space quasar-line emitter two-point correlation function $\xi_{\rm{QG}}$ parameterized as
\begin{equation}
\xi_{\rm{QG}} = \left(r/r_0^{\rm{QG}}\right)^{-\gamma_{\rm{QG}}} \ , 
\end{equation}
where $r=\sqrt{R^2+Z^2}$, and $r_0^{\rm{QG}}$ is the cross-correlation length, and $\gamma_{\rm{QG}}$ is its power-law slope. We perform an MCMC fit to the measured cross-correlation function using \texttt{emcee} \citep{emcee}. Since the quasar-galaxy cross-correlation function measured from [\ion{O}{3}] emitters is consistent with that of [\ion{C}{2}] emitters, we fit all data together. When fixing the slope at $\gamma_{\rm{QG}} = 2.0$ to enable a comparison to other studies \citep{Shen10,Eilers24}, we derive $r_0^{\rm{QG}} = 8.68^{+0.51}_{-0.55}~h^{-1}~{\rm cMpc}$ for ASPIRE quasars with an average redshift of $\langle z\rangle=6.6645$. The fitting result is shown in Figure~\ref{fig:ccf}. For comparison, \cite{Eilers24} derived $r_0^{\rm{QG}} = 9.1^{+0.5}_{-0.6}~h^{-1}~{\rm cMpc}$ using four EIGER quasars at $\langle z\rangle=6.25$. 

If we do not freeze the slope during the fitting, we obtain $\gamma_{\rm{QG}} = 1.68^{+0.18}_{-0.20}$ and $r_0^{\rm{QG}} = 12.20^{+3.47}_{-2.30}~h^{-1}~{\rm cMpc}$. Nevertheless, there is a strong degeneracy between  $\gamma_{\rm{QG}}$ and $r_0^{\rm{QG}}$ as shown in Figure \ref{fig:corner}. Therefore, we use the value from the fixed $\gamma_{\rm{QG}}$ in the following analyses, similar to previous studies \citep{Eilers24}.

\subsection{Quasar host dark matter halos and duty cycles}

Assuming that quasars and [\ion{O}{3}] emitters trace the same underlying dark matter density field \cite{Garcia17}, the quasar auto-correlation $\xi_{\rm{QQ}}$ can be inferred from the galaxy-galaxy auto-correlation $\xi_{\rm{GG}}$ and the quasar-galaxy cross-correlation $\xi_{\rm{QG}}$ according to 
\begin{equation}
\xi_{\rm{QG}} = \sqrt{\xi_{\rm{QQ}}\xi_{\rm{GG}}}
\end{equation}
After applying the $r_0^{\rm{GG}}$ of [\ion{O}{3}] emitter ($r_0^{\rm{GG}}=4.78^{+0.50}_{-0.55}\,\rm{h}^{-1}\,\rm{cMpc}$) measured directly from ASPIRE galaxies (Huang et al. in prep), we derive an quasar auto-correlation length of $r_0^{\rm{QQ}}=15.76^{+2.48}_{-2.70}~h^{-1}~{\rm cMpc}$. 
To the first order, we can constrain the minimum halo mass $M_{\rm{halo, min}}$ by linking $r_0^{\rm{QQ}}$ to the auto-correlation length of dark matter halos with $M_{\rm halo}>M_{\rm{halo, min}}$. Using the \texttt{halomod} \citep{Murray13,Murray21} tool\footnote{\url{https://github.com/halomod/halomod}} with an assumption of a step-function Halo Occupation Distribution (HOD), 
we estimate that $\rm log(M_{\rm{halo, min}}/M_\odot)= 12.27^{+0.21}_{-0.26}$ with a cumulative abundance of dark matter halos $n_{\rm{halo,min}}= {\rm 54.6^{+510.0}_{-48.5}\, cGpc^{-3}}$ with $M>M_{\rm{halo, min}}$. 
This simple calculation suggests that the most distant luminous quasars, on average, reside in the most massive dark matter halos at their respective cosmic times, consistent with expectations from theoretical models \citep{Pizzati24a}.

The spatial density of ASPIRE quasars (i.e., $6.5<z<7$ and $M_{1450}>-25$) are estimated to be $n_{\rm Q}=\rm 0.76~cGpc^{-3}$ based on the luminosity function from \cite{Wang19b} and \cite{Matsuoka18c}. Together with $n_{\rm{halo,min}}$, we can then infer quasars' duty cycle ($f_{\rm duty}$) and UV-luminous life time ($t_Q$) using 
\begin{equation}
n_{\rm Q} = f_{\rm duty} \times n_{\rm halo}
\end{equation}
, where $f_{\rm duty}\simeq\frac{t_Q}{t_{\rm H}(z)}$ and $t_{\rm H}(z)$ is the Hubble time at redshift $z$. After plugging in the numbers, we estimate that $t_{\rm Q}=\rm 10^{{7.05}^{+0.95}_{-1.01}}~yr$ and $f_{\rm duty}=1.4^{+11.1}_{-1.3}$\%. 
We note that the minimum halo mass of ASPIRE quasars, $M_{\rm halo, min}$, lies at the extreme high-mass end of the dark matter halo distribution at these redshifts. Because the halo mass function is extremely steep in this regime, the corresponding number density, $n_{\rm halo, min}$, is highly sensitive to even small changes in $M_{\rm halo, min}$. As a result, estimates of the quasar duty cycle and lifetime are subject to substantial uncertainties. This sensitivity underscores the challenge of constraining the physical properties of high-redshift quasars using halo abundance matching, particularly for the rarest, most massive systems.
The short UV-luminous quasar lifetime and small duty cycle inferred for the ASPIRE quasars suggest that most of the black hole mass growth likely took place during an obscured, non-UV-luminous phase. This scenario would help mitigate the otherwise severe tension associated with forming billion-solar-mass black holes from light seeds by such early cosmic times \citep{Davies19,Wang21a}.

\begin{figure}
\centering
\includegraphics[width=0.49\textwidth]{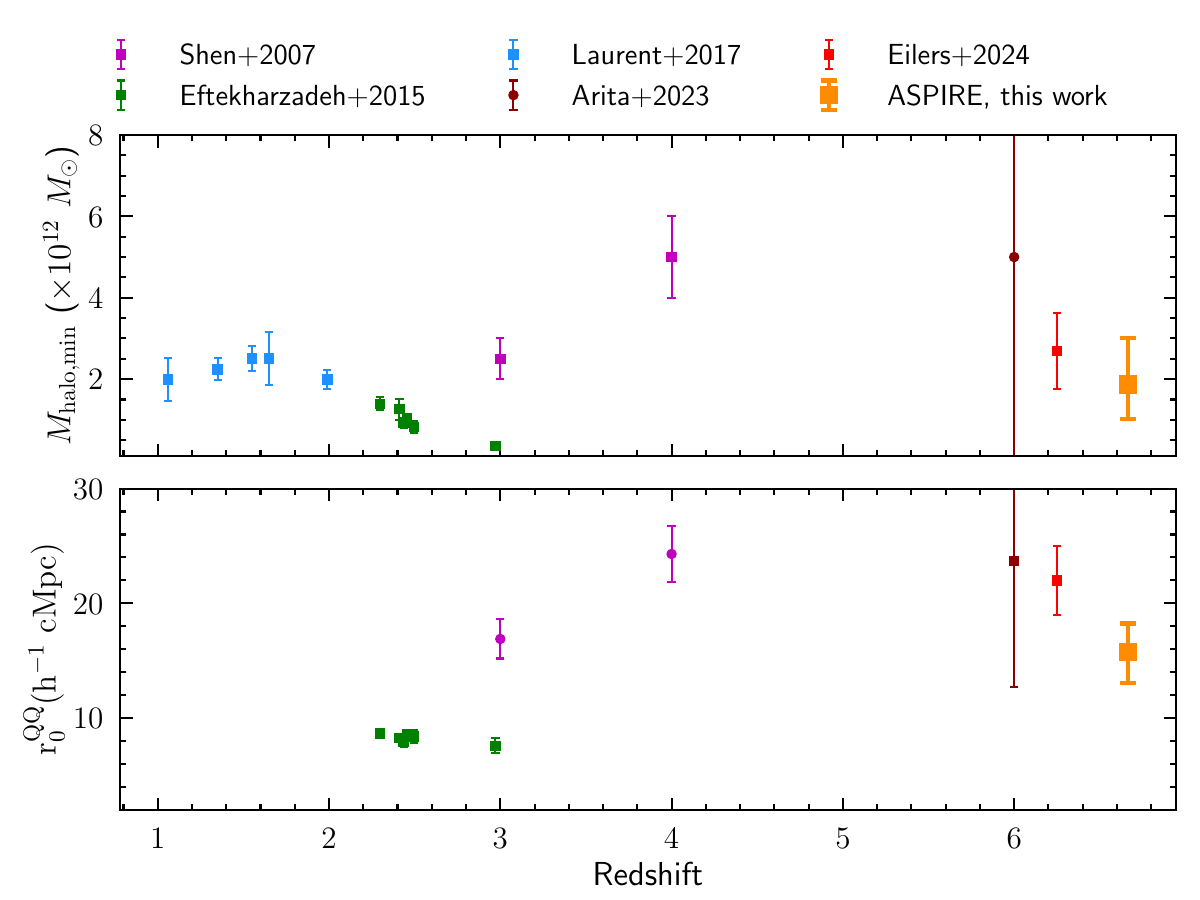}
\caption{{\bf Redshift evolution of correlation length and dark matter halo mass of quasars.}
This work provides the first correlation length and dark matter halo mass measurements of quasars at $z>6.5$. This work indicates that the characteristic dark matter halo mass of quasars does not evolve significantly from the local universe to redshift $z \sim 7$, remaining at a few times $10^{12}~M_\odot$. 
\label{fig:mhalo}}
\end{figure}

We also compare the clustering of galaxies in these quasar fields with the predictions from the \textsc{Trinity} model \citep{Zhang23}. \textsc{Trinity} infers the empirical mass and growth rate connections between dark matter halos, galaxies, and SMBHs, which match a compilation of galaxy data from $z=0-13$ and SMBH data from $z=0-7$ by construction. To make predictions of galaxy clustering around $z\sim 6.5$ quasars, we first generate mock halo--galaxy--SMBH catalogs based on the best-fitting \textsc{Trinity} model and the \textit{Uchuu} simulation \citep[][box size: 2 cGpc$/h$]{Ishiyama21}. To fully utilize the statistical power provided by this large catalog, we use \emph{all} the AGNs in the catalog to calculate the weighted AGN--galaxy cross-correlation function, where each AGN is weighted to reflect their similarity to the ASPIRE quasars. Specifically, the weight $w$ for a simulated AGN with black hole mass $M_\mathrm{BH}$ and bolometric luminosity $L_\mathrm{bol}$ is a sum of its probability given the observed ASPIRE sample, $\{(M_{\mathrm{BH},i}, L_{\mathrm{bol},i})\}_{i=1}^{25}$:

\begin{align}
    w = &\sum_{i=1}^{N=25} P\left[(M_\mathrm{BH}, L_\mathrm{bol}) | (M_{\mathrm{BH},i}, L_{\mathrm{bol},i})\right]\nonumber\\
        =&\sum_{i=1}^{N=25} \frac{1}{2\pi \sigma_{\mathrm{L_\mathrm{bol}},i} \sigma_{M_\mathrm{BH},i}}\nonumber \\
                                &\times \exp{\left[-\frac{(\log M_\mathrm{BH} - \log M_{\mathrm{BH},i})^2}{2\sigma_{M_{\mathrm{BH},i}}^2}\right]}\nonumber\\
                                &\times \exp{\left[-\frac{(\log L_\mathrm{bol} - \log L_{\mathrm{bol},i})^2}{2\sigma_{\mathrm{L_\mathrm{bol}},i}^2}\right]}\ ,
\label{eq:gauss_weights}
\end{align}
where $\sigma_{M_{\mathrm{BH},i}}$ and $\sigma_{\mathrm{L_\mathrm{bol}},i}$ are the measurement uncertainties in SMBH mass and luminosity of the $i$th ASPIRE quasar, respectively. For $\sigma_{M_{\mathrm{BH},i}}$, we also add a scatter of 0.5 dex in quadrature to account for the random scatter in virial estimates of $M_\mathrm{BH}$ \citep{Vestergaard06, Shen13}. To keep the consistency between our predictions and the observations, we applied the $L_\mathrm{\left[OIII\right],4960+5008}$--$M_\mathrm{UV}$ scaling relation from \citet{Matthee24} to calculate the predicted $L_\mathrm{\left[OIII\right],5008}$ for our mock galaxies, assuming $L_\mathrm{\left[OIII\right],5008}:L_\mathrm{\left[OIII\right],4960}=3:1$, and only include galaxies with $L_\mathrm{\left[OIII\right],5008} \geq 10^{42}$ erg/s when calculating the clustering. The \textsc{Trinity} predicted quasar-[\ion{O}{3}] emitter cross-correlation function is shown in Figure \ref{fig:ccf}. The correlation length and slope are $r_0^{\rm{QG}} = 7.9\pm0.4~h^{-1}~{\rm cMpc}$ and $\gamma_{\rm{QG}} = 2.05\pm0.06$ which are consistent with the ASPIRE measurement. The corresponding median halo mass in \textsc{Trinity} is $\rm log(M_{\rm{halo}}/M_\odot)= 12.04{\pm0.22}$, also similar to our measurements. 

\begin{figure*}
\centering
\includegraphics[width=0.98\textwidth]{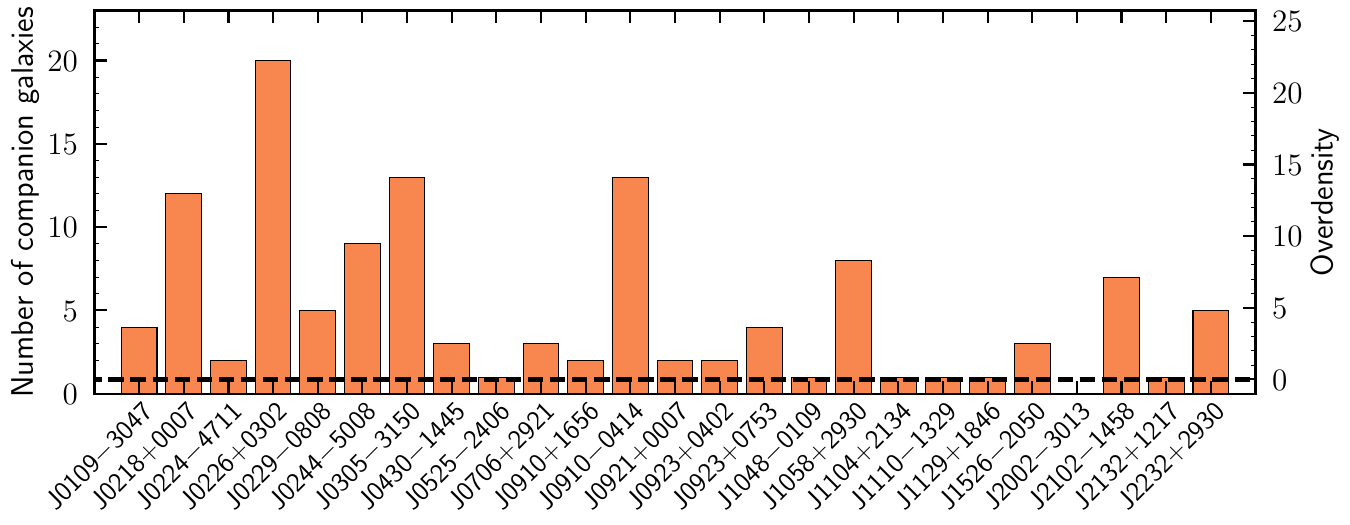}
\caption{{\bf Large variance of the number of quasar companion galaxies.}
 Galaxy numbers are determined by counting [\ion{O}{3}] emitters within $\Delta |v_{\rm los}| < 1000~\mathrm{km~s^{-1}}$ relative to the central quasars. The number of companion galaxies varies significantly from field to field, ranging from zero (lower than expected) to 20 (corresponding to $\delta_{\rm gal} \simeq 22$).
\label{fig:ngal}}
\end{figure*}

As shown in Figure~8, both the ASPIRE measurements and the \textsc{Trinity} predictions exhibit an excess in the quasar-galaxy cross-correlation function, $\xi_{\rm{QG}}$, relative to the power-law fit at $R \lesssim 0.2~h^{-1}~\mathrm{cMpc}$. This excess indicates the presence of the so-called \textit{one-halo term} in HOD models, where an increased number of companion galaxies are found within quasar host halos or sub-halos. 
Such an enhancement is consistent with quasars residing in massive dark matter halos that can host multiple galaxies in close proximity, or with a high satellite galaxy fraction around quasars. Together with the strong [\ion{C}{2}] detections of ASPIRE quasar host galaxies \citep{Wang24b}, the observed small-scale excess from both [\ion{O}{3}] and [\ion{C}{2}] emitters, both indicative of galaxies with active star formation, suggests that the quasar environment is undergoing vigorous star-forming activity. This enhanced activity may be driven by the same gas-rich conditions that fuel quasar accretion, such as filamentary inflows or galaxy mergers. Furthermore, quasar feedback may not yet have suppressed star formation in nearby galaxies at these early cosmic epochs, consistent with the analyses of galaxy properties in quasar vicinity \citep{Champagne24a, Champagne24b}.

In Figure~\ref{fig:mhalo}, we compare the derived $M_{\rm halo, min}$ and $r_0^{\rm QQ}$ with measurements of quasars at lower redshifts from the literature \citep{Shen07, Eftekharzadeh15, Laurent17, Arita23, Eilers24}. Our results suggest that the characteristic dark matter halo mass of quasars does not evolve significantly with redshift up to $z \sim 7$, remaining at a few times $10^{12}~M_\odot$. This indicates that quasars can be triggered in halos of similar mass across cosmic time, consistent with previous findings at lower redshifts. 
This is strikingly close to the mass scale of $M_{\rm halo} \sim 10^{12}~M_\odot$, where galaxy formation is most efficient due to a balance between gas cooling and feedback processes that regulate star formation \citep{Wechsler18}. If these quasars follow the stellar-to-halo mass relation of galaxies \citep{Behroozi19}, our study suggests that these luminous $z \sim 7$ quasars are hosted by galaxies with stellar masses of a few times $10^{10}~M_\odot$, consistent with recent JWST observations of stellar light from quasar hosts \citep[e.g.,][]{Ding23, Yue24}, as well as dynamical mass estimates from ALMA observations \citep[e.g.,][]{Neeleman21, Wang24b}. Since the ASPIRE program also includes deep, high-resolution JWST/NIRCam imaging, the stellar masses of the ASPIRE quasar hosts will soon be available (Yang et al., in prep). 
Combined with the halo masses derived here, this will allow us to directly constrain the stellar-to-halo mass relation for the ASPIRE quasar sample.
Furthermore, we will compare the ASPIRE measurements with cosmological simulations and discuss the constraints on the duty cycle of both quasars and the H$\beta$+[\ion{O}{3}] emitting galaxies in a companion paper (Huang et al., in prep).

\section{Quasar Environment}\label{sec:environment}
\subsection{Diverse quasar environments}

\citet{Habouzit19} investigated the number of companion galaxies around high-redshift SMBHs using the {\tt Horizon-AGN} simulation. They found that, on average, SMBHs are surrounded by more companion galaxies than blank fields, particularly when observations are sensitive to faint or low-mass galaxies (i.e., $M_\ast > 10^8~M_\odot$; see their Figures~3 and 4). However, when only the most massive galaxies are detectable (e.g., $M_\ast > 10^{9\text{--}9.5}~M_\odot$) or when the field of view is restricted, the observed number of companions can vary significantly because of cosmic variance. The differing observational sensitivities, limited survey volumes, limited number of studied quasar fields and the high levels of incompleteness and contamination associated with photometric galaxy selections in previous studies have led to inconclusive results regarding whether the observed diversity in quasar environments arises from cosmic variance, selection bias or reflects intrinsic differences in their underlying environments.

In contrast to earlier studies \citep[e.g.,][]{Willott05,Kim09}, the ASPIRE program provides substantially improved sensitivity to star-forming galaxies, reaching stellar masses as low as $M_\ast \gtrsim 10^{7\text{--}8}~M_\odot$ \citep{Champagne24b}. The WFSS observations also probe much larger comoving volumes along the line of sight, enabling a more complete census of galaxies both in quasar environments and at other redshifts. Furthermore, the large number of independent sightlines observed by ASPIRE significantly reduces the impact of field-to-field cosmic variance, which has been a major limitation in previous surveys. As a result, the number of companion galaxies identified in this study is far less affected by selection biases and cosmic variance and can be compared with theoretical simulations in a more robust and physically meaningful manner. 

In Figure~\ref{fig:ngal}, we show the number of quasar companion [\ion{O}{3}] emitting galaxies and the galaxy overdensity, defined as  $\delta_{\rm gal} = \frac{n_{\rm companion}}{\bar{n}} - 1$, in each of the ASPIRE fields. The galaxy overdensity $\delta_{\rm gal, [OIII]}$ is measured using galaxies within NIRCam module A (with $r_{\rm eff} \lesssim 3~{\rm cMpc}$, or $V \sim 500~{\rm cMpc}^3$ assuming a projected length of $\rm \Delta |v_{\rm los}|\le1000~km~s^{-1}$), where we have uniform depth and continuous coverage. Figure~\ref{fig:ngal} demonstrates that the number of quasar companion galaxies ranges from zero to 20, corresponding to overdensities from zero to $\delta_{\rm gal, [OIII]} \simeq 22$. Specifically, there are seven quasar fields with $\delta_{\rm gal, [OIII]} > 5$, six quasar fields consistent with the number of galaxies in blank fields, one field without any companion galaxy, and the remaining fields with $1 < \delta_{\rm gal, [OIII]} < 5$. 

\begin{figure}
\centering
\includegraphics[width=0.49\textwidth]{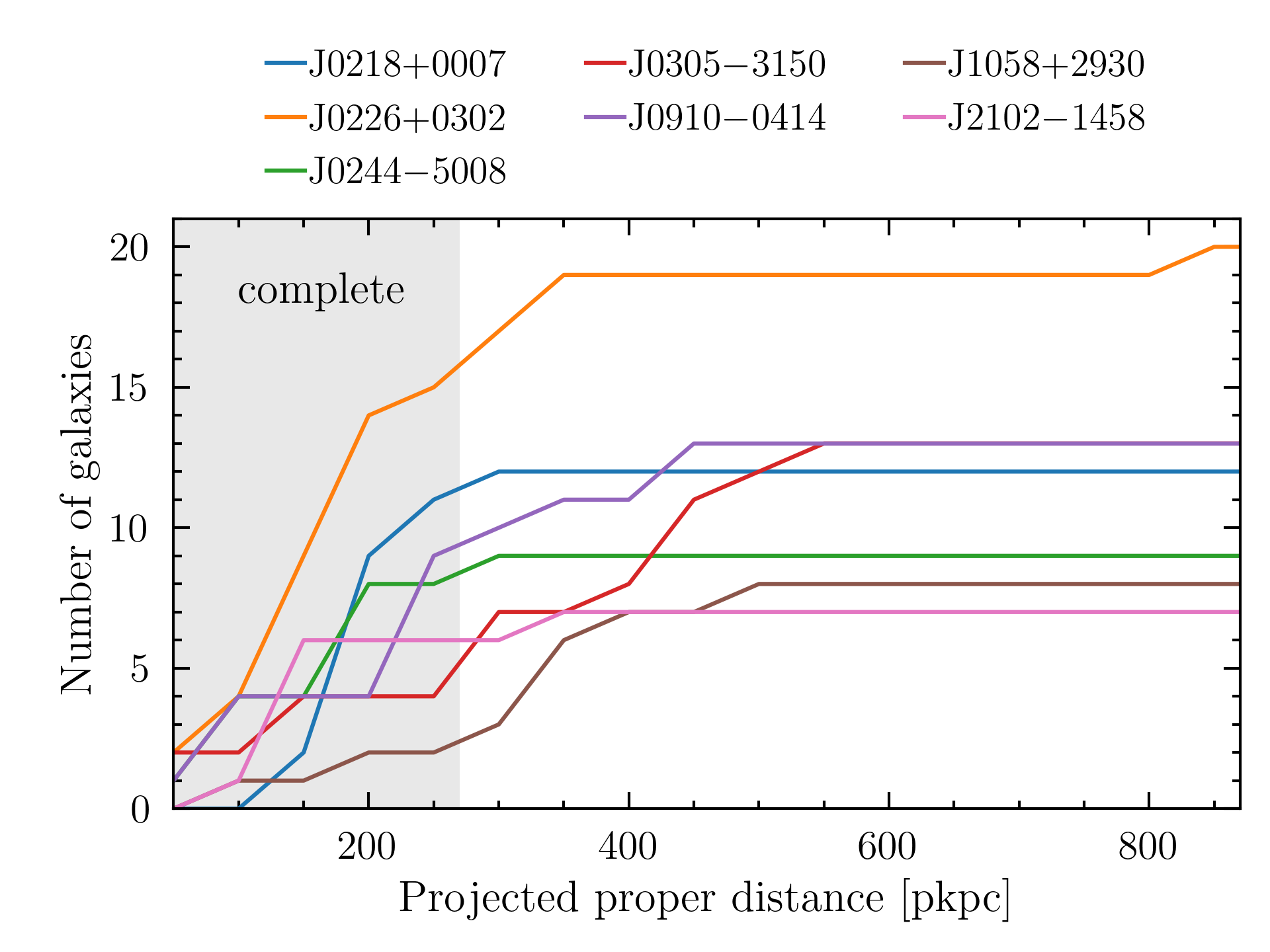}
\caption{{\bf The cumulative galaxy number as a function of projected distance in the seven protocluster field.}
The cumulative number profile is different from field to field. Note that the survey is only complete with $R\lesssim300~{\rm pkpc}$. 
\label{fig:cum_num}}
\end{figure}

\begin{figure*}
\raggedright
\includegraphics[trim=0 0 3cm 0, clip, height=5.4cm]{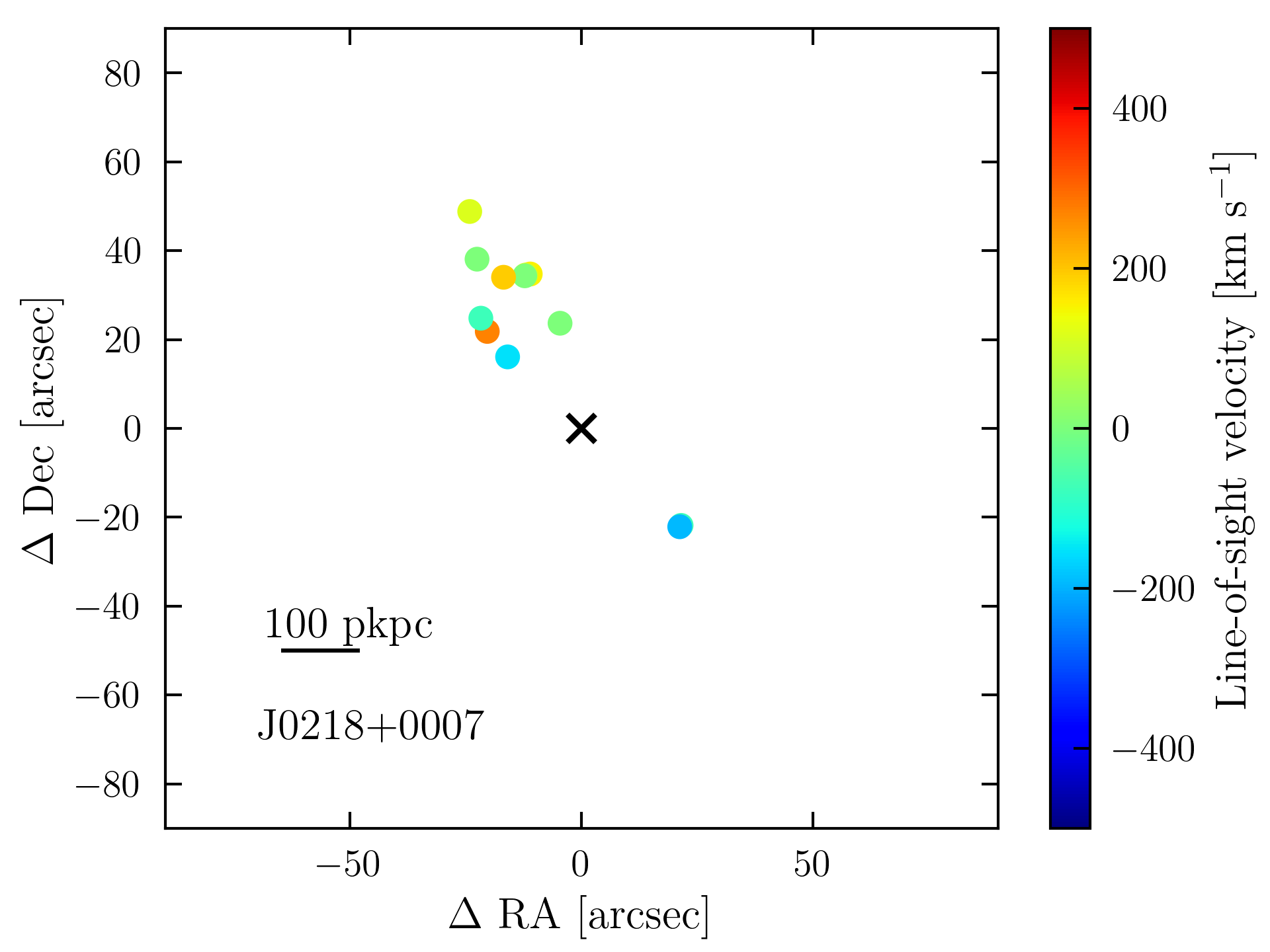}
\includegraphics[trim=0 0 3cm 0, clip, height=5.4cm]{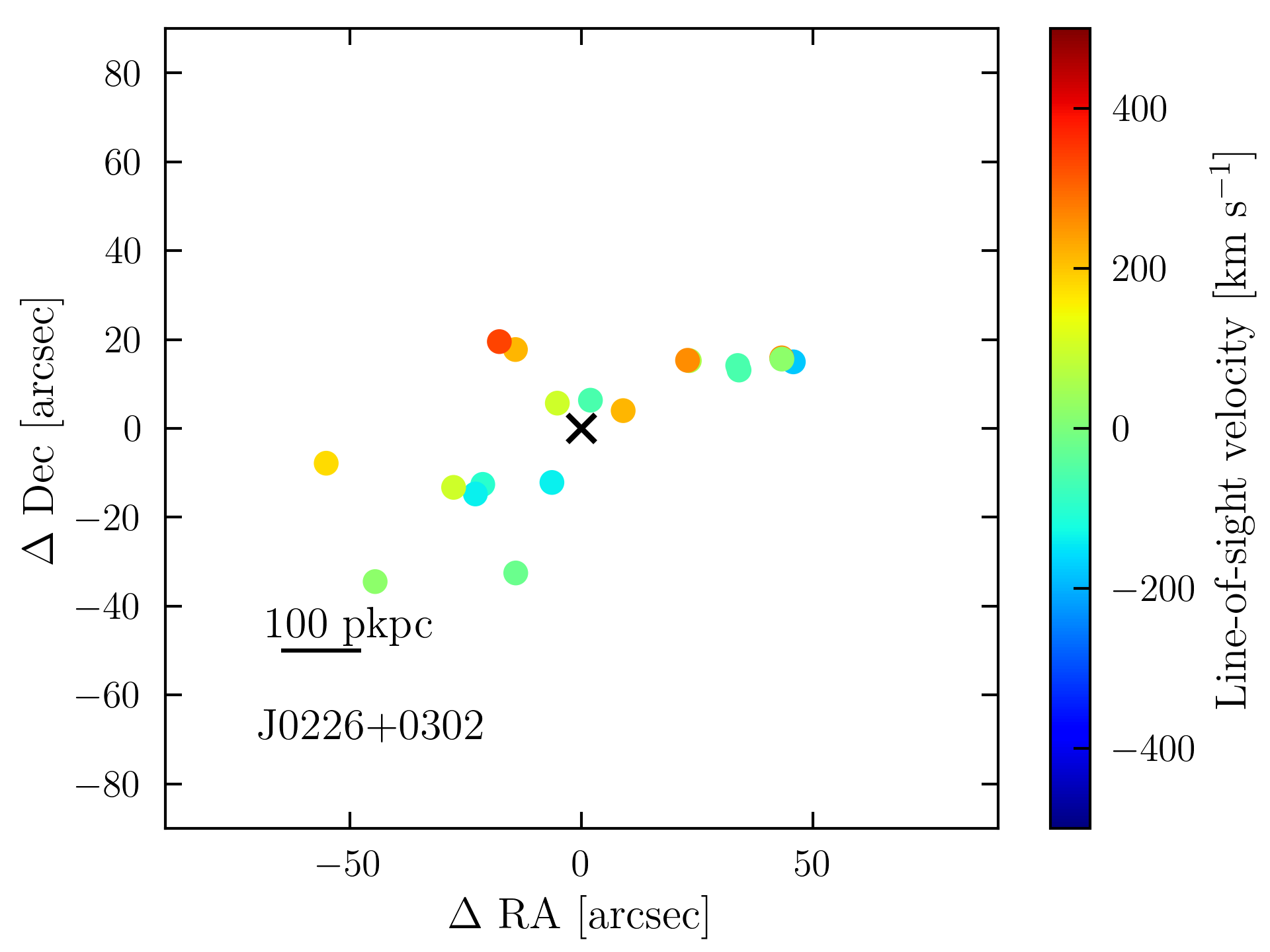}
\includegraphics[trim=0 0 3cm 0, clip, height=5.4cm]{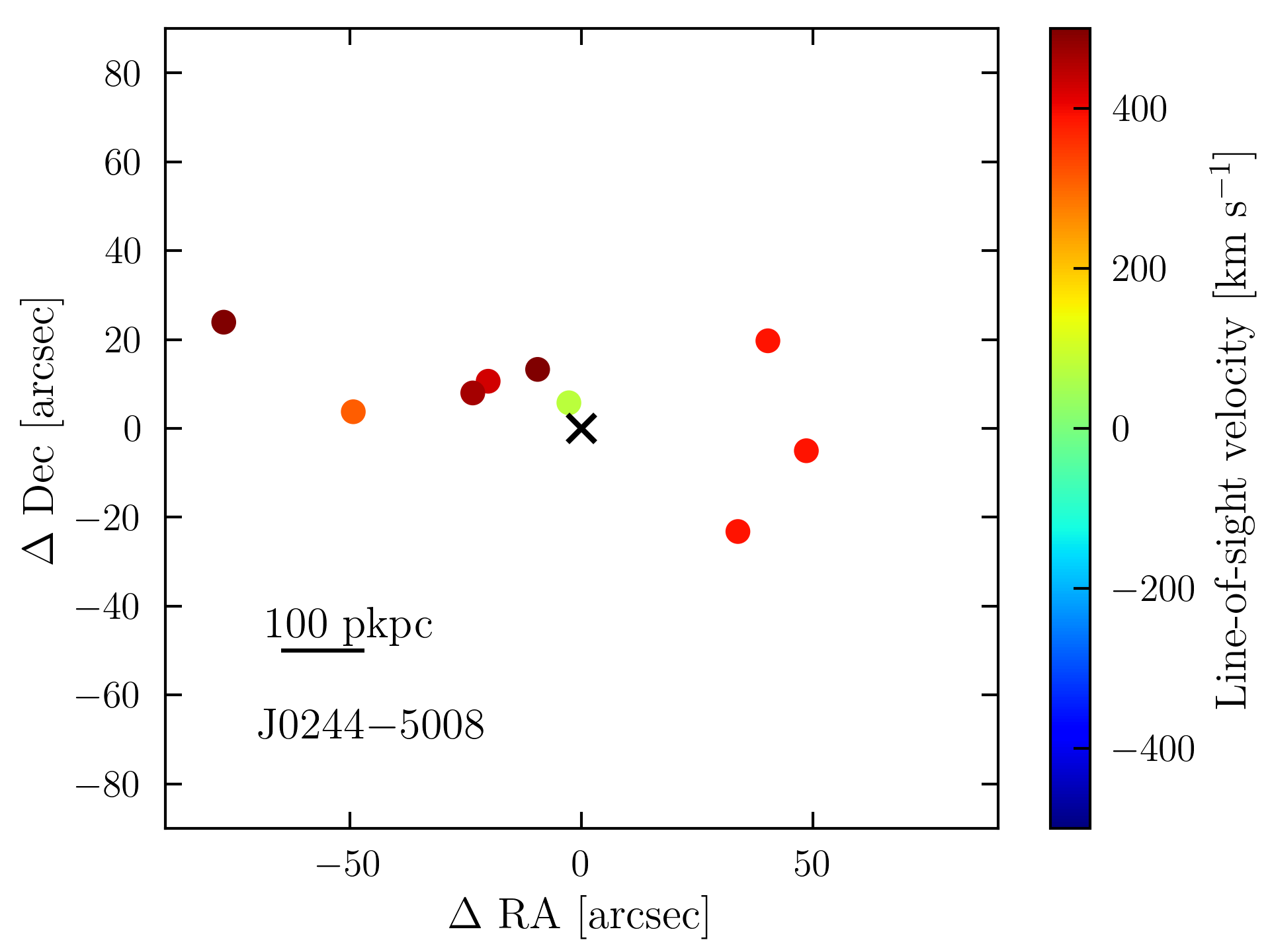}
\includegraphics[trim=0 0 3cm 0, clip, height=5.4cm]{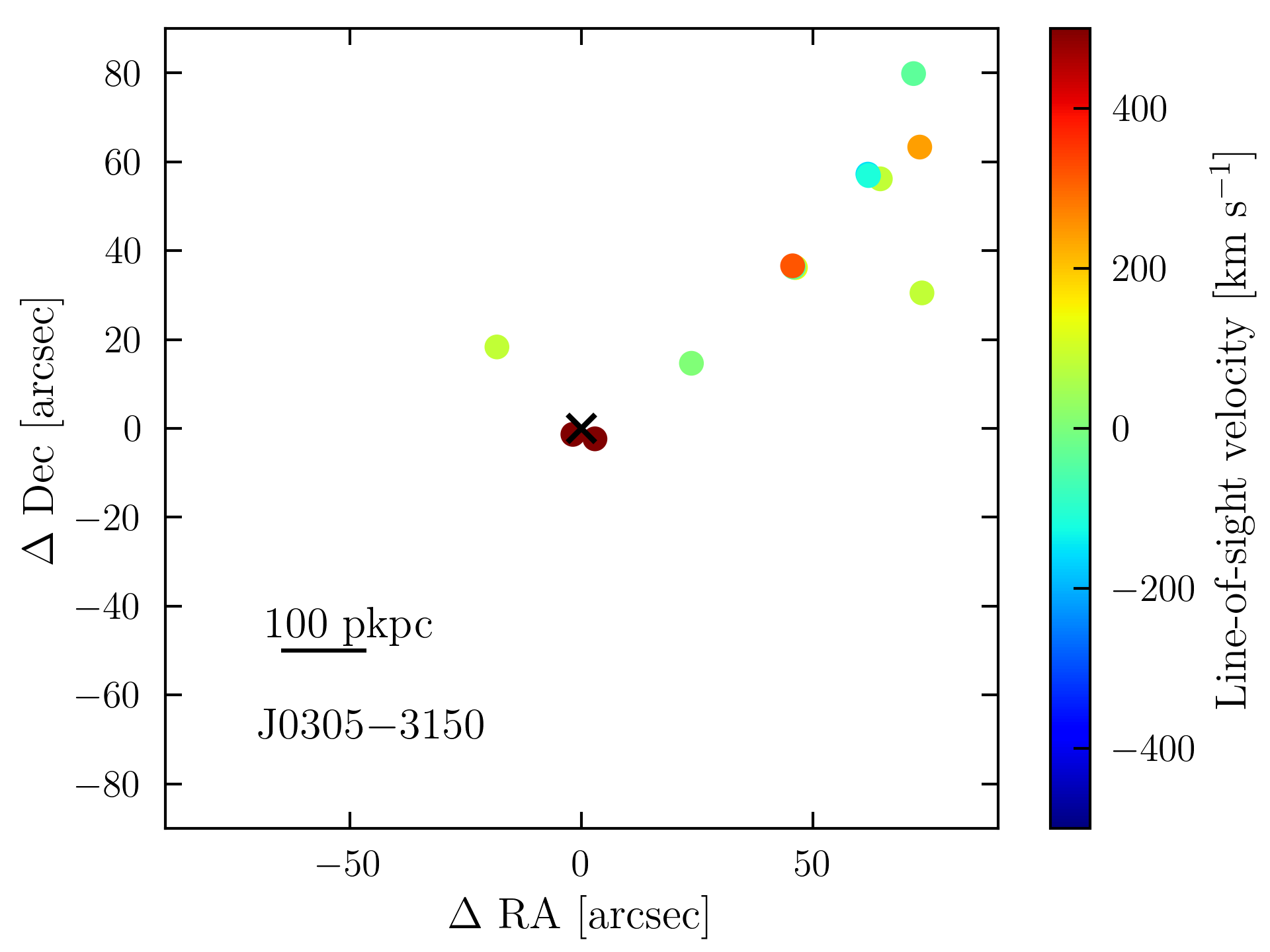}
\includegraphics[trim=0 0 3cm 0, clip, height=5.4cm]{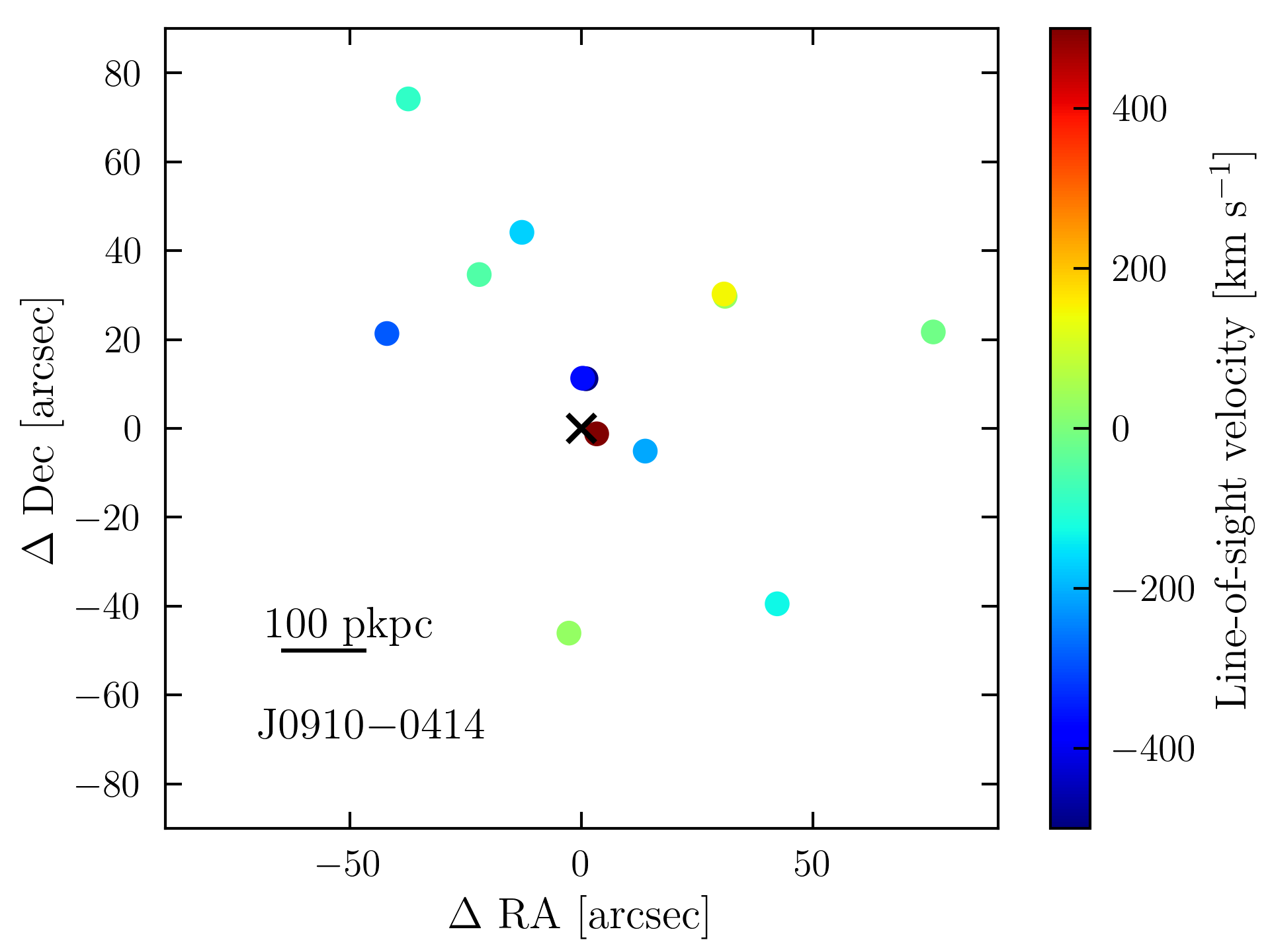}
\includegraphics[trim=0 0 3cm 0, clip, height=5.4cm]{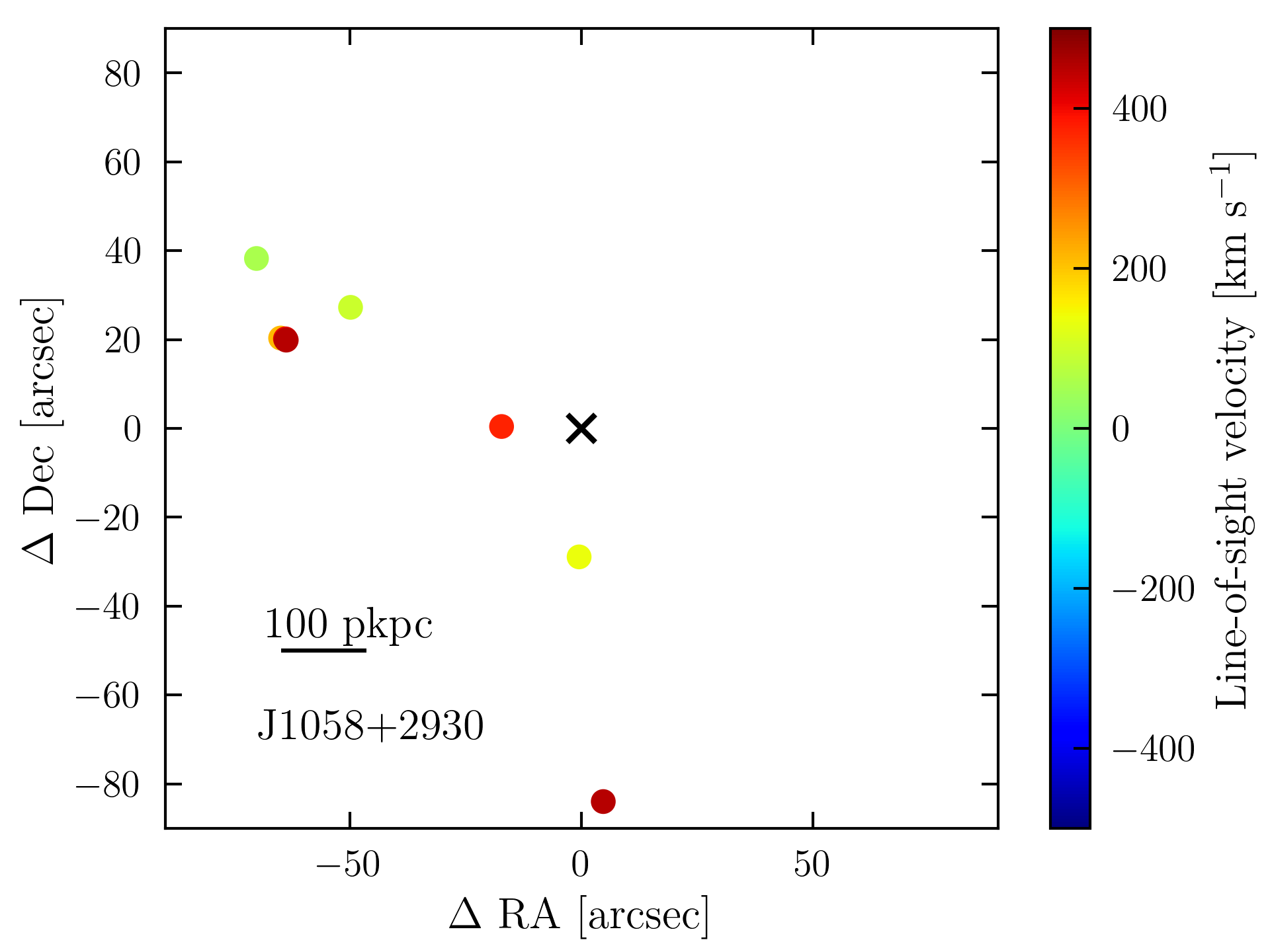}
\includegraphics[height=5.4cm]{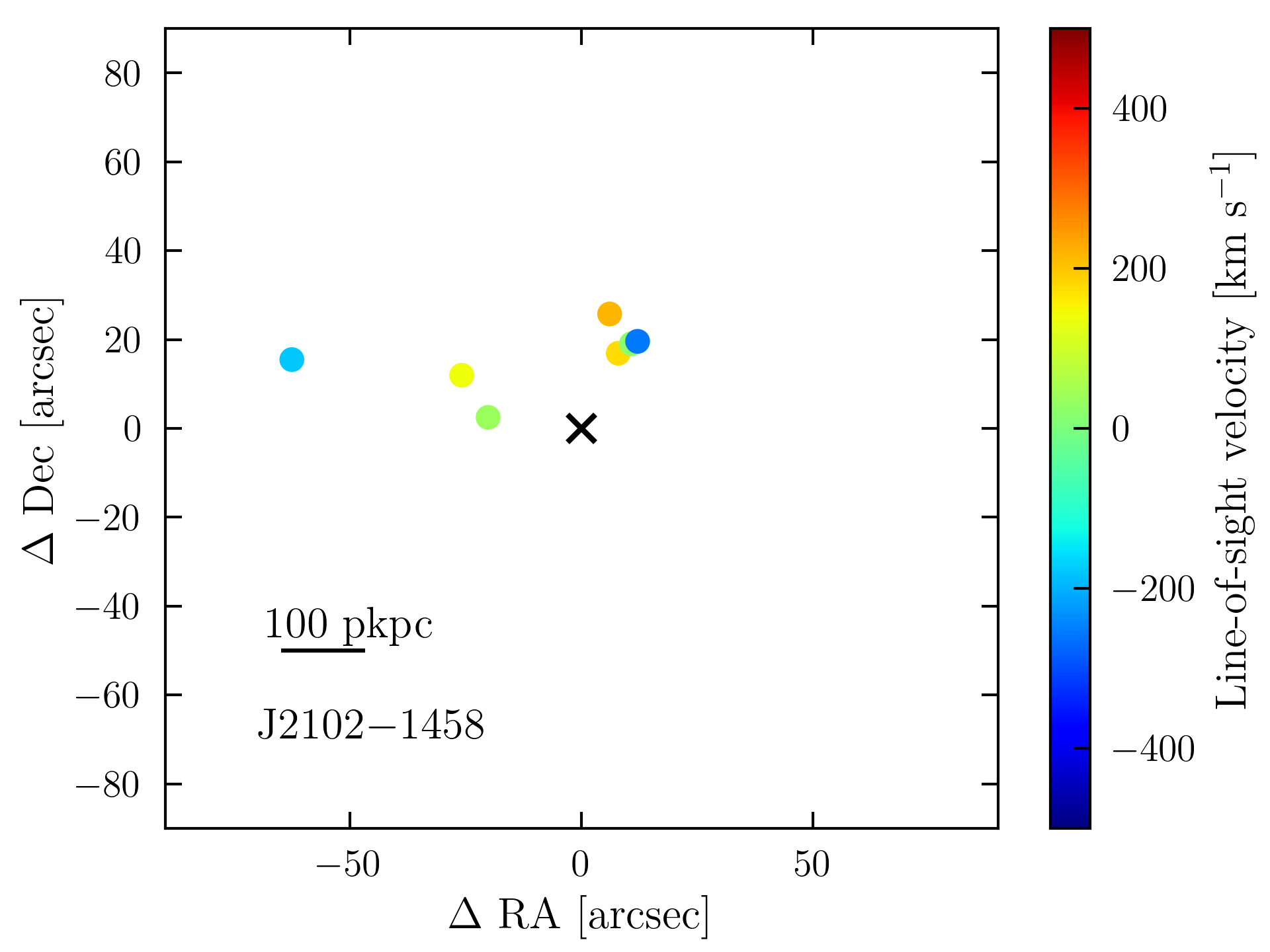}
\caption{{\bf The spatial distribution of [\ion{O}{3}] emitters in the most overdense ASPIRE quasar fields.}
We identified seven quasars inhabit significant galaxy overdensities ($\delta_{\rm gal}>5$ within $V\sim500\, {\rm cMpc^3}$). The black crosses mark the positions of the central quasars, while the colored circles indicate the locations of [\ion{O}{3}] emitters relative to them. The symbols are color-coded by the line-of-sight velocity of the galaxies with respect to the central quasars.
\label{fig:protocluster}}
\end{figure*}

As discussed in \citet{Wang23}, the median number of companion galaxies around SMBHs at $z \sim 6$ with $M_\ast \geq 10^8~M_\odot$ is approximately four in both the \texttt{SIMBA} \citep{Dave19} and \texttt{EAGLE} \citep{Schaye15} simulations, when matched to the depth and survey volume of the ASPIRE program. This prediction is consistent with the average number of companion galaxies observed by ASPIRE ($\bar{n}_{\rm [OIII]} = 4.8$). Moreover, the simulated number of companion galaxies within a survey volume comparable to that of ASPIRE exhibits substantial field-to-field variation, ranging from 2 to 66 when considering all SMBHs with $M_{\rm BH} \gtrsim 10^{7.5}~M_\odot$ across multiple simulation suites, including \texttt{Horizon-AGN} \citep{Dubois14}, \texttt{Illustris} \citep{Vogelsberger14}, \texttt{TNG100/300} \citep{Pillepich18}, \texttt{EAGLE}, \texttt{SIMBA}, \texttt{Astrid} \citep{Bird22,Ni22}, and \texttt{BlueTides}. This broad range is consistent with the large field-to-field variance observed in ASPIRE, as shown in Figure~\ref{fig:ngal}.
Taken together, these advantages allow us to interpret the observed variations in galaxy overdensity (Figure~\ref{fig:ngal}) as evidence of intrinsically diverse quasar environments at high redshift, rather than selection bias or cosmic variance. 
This finding highlights the importance of statistical completeness in environmental studies of rare populations. Future investigations aiming to characterize the environments of luminous quasars, or other rare systems should therefore target a large number of sightlines to minimize the influence of cosmic variance and selection bias.

Interestingly, we find a clear correspondence between the overdensity of [\ion{O}{3}] emitters and the presence of [\ion{C}{2}]-emitting companion galaxies. 
{Specifically, four of the seven fields with significant [\ion{O}{3}] overdensities ($\delta_{\rm gal,[OIII]} > 5$) contain [\ion{C}{2}] companions, compared to only two of the seven fields with low overdensities ($\delta_{\rm gal,[OIII]} \le 1$). While the small number of fields limits the statistical significance, the relative fractions (57\% versus 29\%) point to a consistent trend in which [\ion{C}{2}]-emitting galaxies preferentially reside in fields that exhibit enhanced [\ion{O}{3}] emitter populations. This supports a connection between the large-scale environments traced by the two line-emitting galaxy populations.}
Given that [\ion{C}{2}] emission arises from the neutral interstellar medium (ISM) in star-forming regions and is a good tracer of dust obscured star formation, whereas [\ion{O}{3}] emission primarily traces galaxies with intense star formation and highly ionized gas in less dusty environment, the observed correspondence between these two populations suggests that they both trace the same underlying mass density field. In other words, regions with strong [\ion{O}{3}] overdensities are also likely to host [\ion{C}{2}]-emitting galaxies, and vice versa. This finding indicates that [\ion{O}{3}]- and [\ion{C}{2}]-emitting galaxies jointly map the large-scale structure surrounding high-redshift quasars, providing complementary perspectives on the physical conditions within early overdense regions. 
The detection of both populations in the vicinity of high-redshift quasars therefore demonstrates that multiple modes of star formation, both dusty and relatively unobscured, coexist within the same large-scale structures. Such coexistence implies that the environments of early quasars were not uniform but instead hosted a diversity of interstellar conditions and star-formation processes. This complexity underscores the importance of multi-line studies in capturing the full picture of galaxy formation and feedback in the early Universe.

\begin{figure}
\centering
\includegraphics[width=0.49\textwidth]{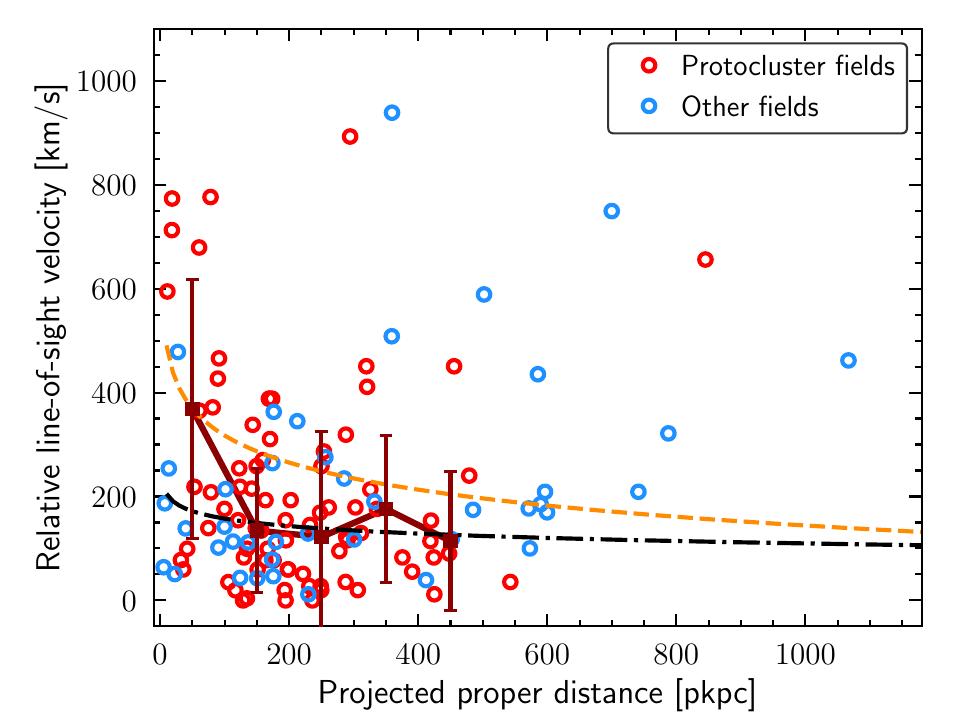}
\caption{{\bf Line-of-sight velocity of [\ion{O}{3}] emitters relative to the central quasars.}
Galaxies in the protocluster fields are shown in red, while those in the other ASPIRE quasar fields are shown in blue. 
{The dark red data points and solid line represent the median line-of-sight velocity of the protocluster fields in five velocity bins (in space of 100 kpc). There are 16, 31, 17, 9, and 7 galaxies in each bins. }
For comparison with cosmological simulations, the orange dashed and black dashed--dotted lines indicate the mean line-of-sight velocity distributions in extremely massive ($M_{\rm halo} > 5 \times 10^{12}~M_\odot$) and moderately massive ($M_{\rm halo} \simeq 5 \times 10^{11}$-$10^{12}~M_\odot$) dark matter halos, respectively, as reported by \citet{Costa24}.
\label{fig:velocity}}
\end{figure}

\subsection{Distribution and kinematics of companion galaxies}
To take a closer look at the structures of the seven fields with $\delta_{\rm gal, [OIII]} > 5$, we show the cumulative number of galaxies as a function of projected distance from the central quasars in these fields in Figure~\ref{fig:cum_num} and the spatial distribution of companion galaxies in Figure \ref{fig:protocluster}. In this work, we define these structures (i.e., regions with $\delta_{\rm gal, [OIII]} > 5$ within $V \sim 500,{\rm cMpc}^3$) as {protocluster candidates}. 
J0226+0302, the most overdense field ($\delta_{\rm gal, [OIII]} > 20$), exhibits the steepest rise within the central $R \lesssim 300~{\rm pkpc}$, where the galaxy survey is complete for sources with $L_{\rm [OIII], 5008\text{\AA}} > 10^{42}~{\rm erg~s^{-1}}$. A more detailed characterization of this structure, along with new JWST mosaic observations, will be presented in future works. J0218+0007 has a similar $\delta_{\rm gal, [OIII]}$ to that of J0305$-$3150 and J0910$-$0414, but shows a more rapid increase in galaxy counts, indicating that the structures traced by J0305$-$3150 and J0910$-$0414 are more spatially extended than that of the structure traced by J0218+0007. This is consistent with the fact that J0305$-$3150 extends well beyond the NIRCam field of view, and that the quasar does not reside in the densest region of the structure \citep{Champagne24a,Champagne24b}.

Interestingly, the companion galaxies are not uniformly distributed but instead appear more consistent with filamentary structures commonly seen in cosmological simulations \citep{Costa24} in all seven protocluster fields. To investigate whether any coherent motions are present, we color-coded the symbols by the line-of-sight velocity, $v_{\rm los} = \frac{z_{\rm gal} - z_{\rm quasar}}{1 + z_{\rm quasar}}c$, of each galaxy relative to its central quasar in Figure \ref{fig:protocluster}. We find no clear evidence of velocity gradients or coherent motions within individual structures. This is similar to what we have seen in cosmological simulations \citep{Costa24} and suggesting that we are observing unvirialized filamentary structures in the early Universe. 
To further study the kinematics of galaxies in these environments, we plot their relative line-of-sight velocities ($|v_{\rm los}|$) as a function of projected distance from the central quasars in Figure \ref{fig:velocity}. 
{In this figure, we also show the median $|v_{\rm los}|$ in seven protocluster fields across five bins. We find that, on average, companion galaxies within a projected distance of $\lesssim 100$ kpc from the quasars exhibit a steep rise in $v_{\rm los}$ in these fields, albeit with small number statistics. This is consistent with the increased velocities of galaxies within the virial radius of massive halos (i.e., the virial radius of a $10^{13}~M_\odot$ halo at $z\sim6.6$ is approximately $\sim100$ kpc). 
We also compare our results with the cosmological simulations by \cite{Costa24}, which studied six extremely massive dark matter halos ($M_{\rm halo}>5\times10^{12}~M_\odot$) and six moderately massive dark matter halos ($M_{\rm halo}\simeq6\times10^{11}-1\times10^{12}~M_\odot$) at $z\sim6$ using high-resolution numerical simulations. The six massive halos represent the most extreme high-density peaks at these redshifts and serve as signposts of the earliest protoclusters, all of which evolve into galaxy clusters by $z=0$.
\cite{Costa24} found that $\sim10\%$ of the satellite galaxies around the high-mass halos have $|v_{\rm los}|$ values close to or exceeding 800 km s$^{-1}$, whereas the steep rise in $|v_{\rm los}|$ was not observed in the lower-mass halos. In Figure~\ref{fig:velocity}, we show the fitted mean $|v_{\rm los}|$ values for the two halo samples from \cite{Costa24}. The protocluster fields identified in this work exhibit $|v_{\rm los}|$ distributions similar to those of the high-mass halos in \cite{Costa24}, with some galaxies reaching $|v_{\rm los}|$ of $\sim 800$ km s$^{-1}$, indicating that they reside in extremely massive halos.}

\begin{figure}
\centering
\includegraphics[width=0.49\textwidth]{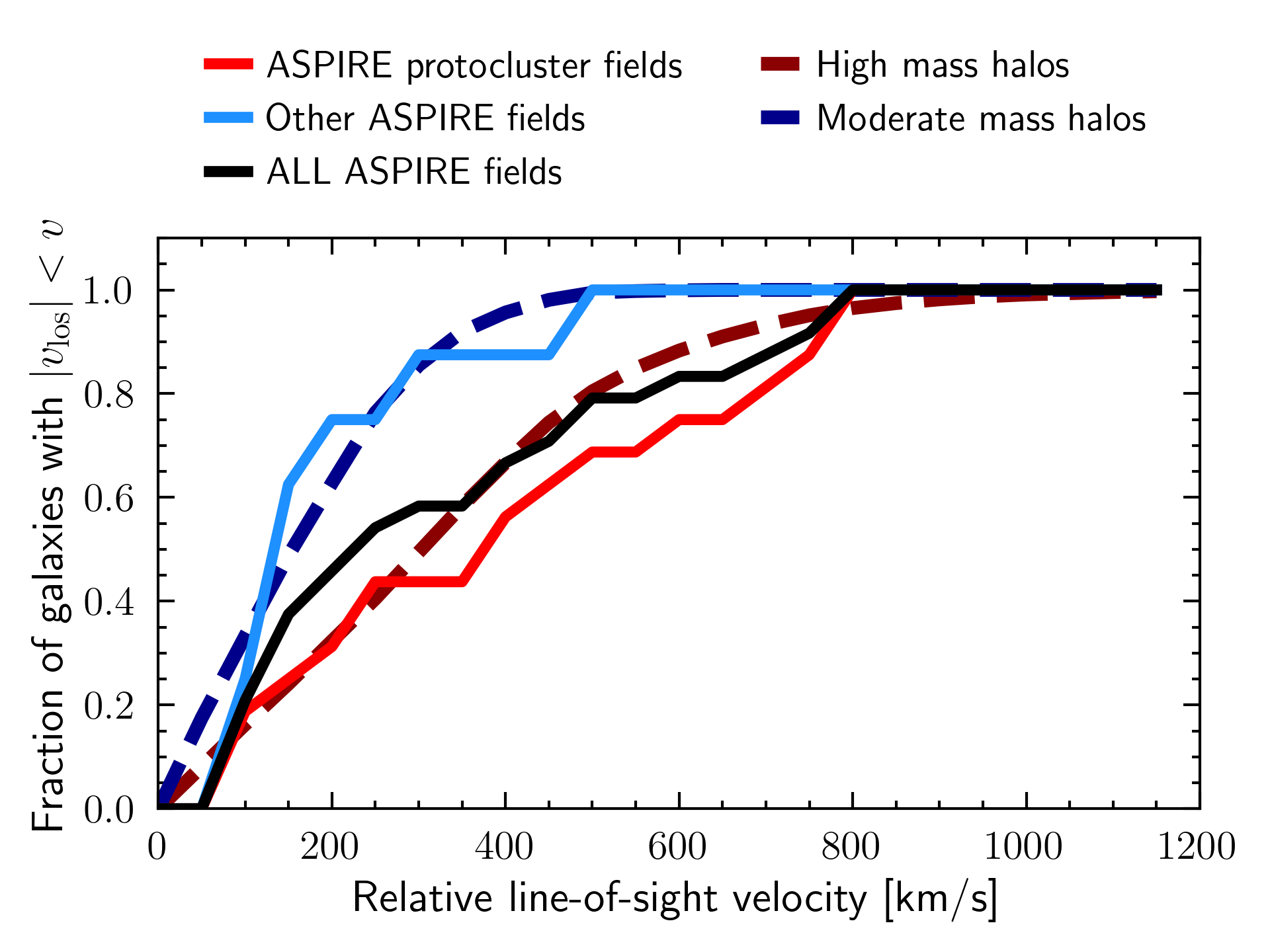}
\caption{{\bf Fraction of galaxies with $v_{\rm los}<v$ within $\rm 100~kpc$ from the central quasars.}
Galaxies in the protocluster fields are shown in red solid line, while those in the other ASPIRE quasar fields are shown in blue solid line. The black solid line represents for all ASPIRE fields. The dark blue and dark red dashed lines are derived from cosmological simulation by \cite{Costa24} for high mass ($M_{\rm halo} > 5 \times 10^{12}~M_\odot$) and moderately massive ($M_{\rm halo} \simeq 5 \times 10^{11}$-$10^{12}~M_\odot$) dark matter halos, respectively.
\label{fig:velocity_cum}}
\end{figure}

To compare our results with simulations more quantitatively, we show the cumulative distribution of $|v_{\rm los}|$, i.e., the fraction of galaxies with $|v_{\rm los}|<v$ within $\rm 100~kpc$ of the central quasars, in Figure~\ref{fig:velocity_cum}. Overall, the $|v_{\rm los}|$ distribution from the ASPIRE program (black solid line) more closely resembles that of high-mass halos (i.e., $M_{\rm halo}>5\times10^{12}~M_\odot$).
When separating the distributions for protocluster and non-protocluster fields, we find that the protocluster fields exhibit an even higher fraction of galaxies with large $|v_{\rm los}|$, suggesting that the seven ASPIRE quasars with $n_{\rm gal}>5$ likely reside in extreme protocluster environments. In contrast, the $|v_{\rm los}|$ distribution for the remaining ASPIRE fields aligns more closely with the moderately massive halo simulations. These comparisons imply that the host dark matter halos of ASPIRE quasars have a broad range, with some residing in very massive dark matter halos tracing extreme protocluster regions.
{However, our current measurements of $|v_{\rm los}|$ rely on a small number of galaxies, and we do not yet have evidence that these structures will be virialized. Therefore, no firm conclusion can yet be drawn about whether they will eventually evolve into massive galaxy clusters by $z=0$.}
Given the significant galaxy overdensities in these fields and the fact that these structures extend to the edges of the JWST/NIRCam field of view, we are conducting follow-up observations to investigate the extent and properties of these filamentary structures. These efforts will also explore the potential role of the central quasars in regulating galaxy formation within these environments \citep[see, e.g.,][]{Champagne24a,Champagne24b}.

\subsection{Implications for the early growth of SMBHs}
{
Our study reveals that a substantial fraction of ASPIRE quasars do not reside in overdense regions, implying a far more diverse range of environments and host halo masses for the first quasars than traditionally assumed. This result stands in contrast to predictions from many cosmological simulations \citep[e.g.,][]{Costa14, DiMatteo17}, which typically require both a massive dark matter halo and a highly overdense large-scale environment in order to grow a seed black hole into a billion-solar-mass quasar under Eddington-limited accretion. The discovery of such diverse quasar environments therefore poses significant challenges to these models. In particular, it suggests that the earliest supermassive black holes must, at least in a substantial subset of cases, follow more rapid or more efficient growth pathways, potentially involving sustained super-Eddington accretion, even for massive seeds, allowing them to reach $\sim10^9\,M_\odot$ within a few hundred million years without requiring exceptionally dense environments. Recent theoretical work provides additional support for this emerging picture. For example, \citet{Fontanot25}, using the GAEA model coupled to the \textit{Planck} Millennium Simulation, found that bright quasars can inhabit a wide range of environments and that secular processes such as disk instabilities may play an important role in triggering early quasar activity. In this scenario, the presence of a significant galaxy overdensity around a quasar is not a necessary condition. }

{
On the other hand, JWST has revealed an unexpectedly large population of faint AGNs at $z>4$ \citep[e.g.,][]{Kocevski23, Greene24, Lin24, Matthee24}, suggesting that our current framework for early black hole growth may be significantly incomplete. Their high abundance indicates that they cannot all reside in very massive dark matter halos \citep{Pizzati25} and that they may represent an alternative pathway toward assembling the $\sim 10^9\,M_\odot$ black holes that power luminous quasars. Notably, a subset of these JWST-detected faint AGNs exhibits strong Balmer breaks and/or Balmer absorption features \citep{Naidu25}, pointing to early black hole growth occurring in dense, gas-rich environments and potentially proceeding at super-Eddington accretion rates \citep{Schleicher13, Coughlin24}. However, it remains unclear whether these faint AGNs are connected to the population of luminous quasars, or whether they could serve as progenitors for a subset of them, particularly for those quasars that do not reside in overdense environments. Taken together, these findings underscore the need to revise theoretical expectations for early black hole growth and highlight the critical importance of conducting comprehensive studies of the environments and clustering properties of both luminous quasars and the faint AGNs uncovered by JWST.
}

\section{Summary}\label{sec:summary}
In this work, we presented an overview of the ASPIRE program, a legacy galaxy redshift survey with the JWST/NIRCam WFSS and ALMA mosaic observations along 25 high-redshift quasar sightlines. This program enabled the discovery of 487 [\ion{O}{3}] emitting galaxies at $5.3 \lesssim z \lesssim 7$, including approximately 120 companion galaxies that are physically associated ($\Delta v_{\rm los} < 1000~\mathrm{km~s^{-1}}$) with the cosmic structures traced by the central luminous quasars. The ALMA observation also enabled the discovery of 17 high-confidence line emitter galaxies with 16 of them are potential [\ion{C}{2}] emitters at the quasar redshifts. 
This large sample of spectroscopically confirmed galaxies allowed us to draw the following conclusions:

\begin{itemize}
    \item On average, luminous $z \sim 7$ quasars are effective tracers of galaxy overdensities, with 18 out of 25 quasars hosting more than one companion [\ion{O}{3}]-emitting galaxy and exhibiting overdensities relative to the expected cosmic average. The number of [\ion{O}{3}]-emitting galaxies at ASPIRE quasar redshifts is 9.4 times higher than that of [\ion{O}{3}]-emitting galaxies at other redshifts. 
    \item The quasar--galaxy cross-correlation functions measured from both [\ion{O}{3}] and [\ion{C}{2}] emitters yield similar results: galaxies are strongly clustered around ASPIRE quasars, suggesting that distant luminous quasars generally reside in massive halos. Detailed clustering analyses based on quasar-[\ion{O}{3}] and quasar-[\ion{C}{2}]  cross-correlation functions indicate that ASPIRE quasars have a auto-correlation length of $r_0^{\rm QQ} = 15.76^{+2.48}_{-2.70}~h^{-1}~{\rm cMpc}$, corresponding to a host halo mass of $\log(M_{\rm halo, min} / M_\odot) = 12.27^{+0.21}_{-0.26}$, the first precise measurement at $z>6.5$. 
    \item The clustering analysis indicates that the UV-luminous duty cycle of ASPIRE quasars is low with $f_{\rm duty}=1.4^{+11.1}_{-1.3}$\% and a UV-luminous lifetime of $t_{\rm Q}=\rm 10^{{7.05}^{+0.95}_{-1.01}}~yr$, reinforces that most of the black hole mass growth likely took place during an obscured UV-faint phase. 
    \item The number of quasar companion [\ion{O}{3}]-emitting galaxies varies significantly from field to field, ranging from zero to 20, corresponding to overdensity of zero to 22. By comparing with cosmological simulations, we conclude that the diverse number of quasar companion galaxies reflects the intrinsically diverse quasar environments rather than being driven by observational biases. 
    \item A clear correspondence between the overdensity of [\ion{O}{3}]-emitting galaxies and the presence of [\ion{C}{2}]-emitting companion galaxies suggests that [\ion{O}{3}]- and [\ion{C}{2}]-emitting galaxies jointly map the large-scale structure surrounding high-redshift quasars and multiple modes of star formation, both dusty and relatively unobscured, coexist within the same large-scale structures.
    \item Seven ASPIRE quasars reside in extremely overdense structures ($\delta_{\rm gal} > 5$ within $V \sim 500\, {\rm cMpc}^3$) and present the ``finger-of-god" effect, consistent with being among the most extreme protoclusters observed in the early Universe. 
\end{itemize}

This study provides the first statistical constraint on the environments and dark matter halo masses of quasars at $z > 6.5$, using a sample of 25 quasar sight lines and leveraging both obscured and unobscured galaxies as large-scale structure tracers. These results represent a major step forward in our understanding of the large-scale environments of early quasars, complementing existing studies focused on individual systems or smaller samples. Together with forthcoming measurements of quasar host galaxy properties and central black hole masses (Yang et al., in prep), the ASPIRE program will establish a critical benchmark for high-redshift quasar studies, spanning spatial scales from the sub-parsec central engine to the several-Mpc-scale cosmic web. In the coming years, as more deep observations from JWST, ALMA, and other facilities (e.g., Euclid, Rubin, and Roman) become available, we expect to gain a clearer and more complete picture of how the earliest luminous quasars form, grow, and evolve. These results will also help to clarify how the emergence of bright quasars connects to the broader population of faint, narrow-line or broad-line AGN candidates recently discovered by JWST at high redshifts, shedding light on the earliest phases of supermassive black hole and galaxy co-evolution.

\begin{acknowledgments}
F.W. acknowledges support from NSF award AST-2513040.
J.B.C.  acknowledges funding from the JWST Arizona/Steward Postdoc in Early galaxies and Reionization (JASPER) Scholar contract at the University of Arizona.
M.H. acknowledge support from the Swiss SNSF Starting Grant(grant no. 218032).
SEIB is supported by the Deutsche Forschungsgemeinschaft (DFG) under Emmy Noether grant number BO 5771/1-1.
J.-T.S. is supported by the Deutsche Forschungsgemeinschaft (DFG, German Research Foundation) - Project number 518006966.
AL acknowledges support from PRIN MUR 2022935STW.
C.M. acknowledges support from Fondecyt Iniciacion grant 11240336 and the ANID BASAL project FB210003.
RAM acknowledges support from the Swiss National Science Foundation (SNSF) through project grant 200020\_207349.
B.T.\ acknowledges support from the European Research Council (ERC) under the European Union's Horizon 2020 research and innovation program (grant agreement number 950533), and from the Excellence Cluster ORIGINS which is funded by the Deutsche Forschungsgemeinschaft (DFG, German Research Foundation) under Germany's Excellence Strategy - EXC 2094 - 390783311.
M.V. gratefully acknowledges financial support from the Independent Research Fund Denmark via grant numbers DFF 8021-00130 and  3103-00146 and from the Carlsberg Foundation (grant CF23-0417).

This work is based on observations made with the NASA/ESA/CSA James Webb Space Telescope. The data were obtained from the Mikulski Archive for Space Telescopes at the Space Telescope Science Institute, which is operated by the Association of Universities for Research in Astronomy, Inc., under NASA contract NAS 5-03127 for JWST. These observations are associated with program \#2078 and can be accessed via \dataset[10.17909/vt74-kd84]{https://doi.org/10.17909/vt74-kd84}.
Support for program \#2078 was provided by NASA through a grant from the Space Telescope Science Institute, which is operated by the Association of Universities for Research in Astronomy, Inc., under NASA contract NAS 5-03127. We acknowledge the strong support provided by the program coordinator Weston Eck and instrument reviewers Norbert Pirzkal and Stephanie La Massa.

This paper makes use of the following ALMA data: ADS/JAO.ALMA\#2022.1.01077.L. ALMA is a partnership of ESO (representing its member states), NSF (USA) and NINS (Japan), together with NRC (Canada), MOST and ASIAA (Taiwan), and KASI (Republic of Korea), in cooperation with the Republic of Chile. The Joint ALMA Observatory is operated by ESO, AUI/NRAO and NAOJ. The National Radio Astronomy Observatory is a facility of the National Science Foundation operated under cooperative agreement by Associated Universities, Inc.
\end{acknowledgments}
\clearpage

%

\vspace{5mm}
\facilities{JWST (NIRCam)}


\software{astropy \citep{Astropy},  
Matplotlib \citep{Matplotlib},
Numpy \citep{Numpy},
Photutils \citep{photutils},
Scipy \citep{Scipy},
Source Extractor \citep{SExtractor}
}


\suppressAffiliationsfalse
\allauthors 



\end{document}